\documentclass[twocolumn,secnumarabic,amssymb, nobibnotes, aps, prd,showkeys,showpacs]{revtex4-1}
\usepackage{amsfonts}
\usepackage{amsthm}
\usepackage{amsmath}
\usepackage{graphicx}
\usepackage{dcolumn}
\usepackage{bm}
\usepackage{euscript}

\begin{document}

\title{The quasiclassical theory of the Dirac equation with a
scalar-vector interaction \\ and its applications in the theory of heavy-light mesons}
\author{V.Yu.~Lazur}\email{lazur@univ.uzhgorod.ua.}
\author{O.K.~Reity}\email{reiti@univ.uzhgorod.ua.}
\author{V.V.~Rubish}\email{vrubish@univ.uzhgorod.ua.}
\affiliation{Department of Theoretical Physics, Uzhgorod National University, \\Voloshyna Street 54, Uzhgorod 88000, Ukraine}

\date{\today}

\begin{abstract}
We construct a relativistic potential quark model of $D$, $D_s$, $B$, and $B_s$ mesons in which the light quark motion
is described by the Dirac equation with a scalar-vector interaction and the heavy quark is considered
a local source of the gluon field. The effective interquark interaction is described by a combination of the
perturbative one-gluon exchange potential $V_{\mathrm{Coul}}(r)=-\xi/r$ and the long-range Lorentz-scalar and Lorentz-vector
linear potentials $S_{\mathrm{l.r.}}(r)=(1-\lambda)(\sigma
r+V_0)$ and $V_{\mathrm{l.r.}}(r)=\lambda(\sigma r+V_0)$, where $0\leqslant\lambda<1/2$. Within the quasiclassical approximation, we obtain simple asymptotic formulas for the energy and mass spectra and for the mean radii of $D$, $D_s$, $B$, and $B_s$ mesons, which ensure a high accuracy of calculations even for states with the radial quantum number $n_r\sim 1$. We show
that the fine structure of P-wave states in heavy-light mesons is primarily sensitive to the choice of two
parameters: the strong-coupling constant $\alpha_s$ and the coefficient $\lambda$ of mixing of the long-range scalar and vector potentials $S_{\mathrm{l.r.}}(r)$ and $V_{\mathrm{l.r.}}(r)$. The quasiclassical formulas for asymptotic coefficients of wave function at zero and infinity are obtained.
\end{abstract}

\maketitle

\section{Introduction}\label{s1}

The heavy-light quark-antiquark ($Q\bar{q}$) systems, being QCD analogues of relativistic hydrogen-like atoms, are ideal objects for investigations and permit very
precise experimental verification of quantum theory results. Theoretical description of the mass spectra and decay probabilities
of such composite objects requires the construction of a consistent theory of bound states, which should be based on
the fundamental principles of local quantum field theory and use its apparatus \cite{Bogolyubov}. However, calculating these
characteristics of composite systems directly in the local quantum field theory is not always possible, because
the only known calculation method in this theory is still based on the perturbation theory, while the nature
of creation of a bound state of interacting particles must undoubtedly be determined by nonperturbative
effects.

When constructing the theory of bound states, the most effective way of leaving the framework of the perturbation theory is the use of dynamic equations. The point is that even if we can construct kernels of
dynamical equations only in the lower orders of the perturbation theory, developing methods for solving
them exactly or approximately (but without using the perturbation theory) allows taking into account nonperturbative
effects of interaction when evaluating observable characteristics of the bound states. In a
nonrelativistic case, such a theory is formulated in the language of the classical potential using the dynamical
Schr$\ddot{\mbox{o}}$dinger equation. But at large bound energies, the corresponding theory must be essentially relativistic.
In this regard, the way to solve this problem was indicated about half a century ago based on using
the dynamical equations in the local quantum field theory, examples of which are the Bethe-Salpether
equation \cite{Salpeter}, the quasipotential equation \cite{Logunov}, and other equations \cite{Macke}.

The Dirac equation with a mixed scalar-vector interaction plays an important role in the contemporary development of the relativistic theory of bound states. It is valuable because it provides an adequate mathematical model for a wide circle of problems in hadronic physics in which it is possible to transit consistently from a two-particle problem to the external field approximation. This equation indicates the
presence of the spin and spin moment for the quark and antiquark, and the problems of describing fine
and superfine structures in the energy spectra of heavy-light ($Q\bar{q}$) mesons, which are the QCD analogues
of hydrogen-like atoms, arise naturally from this equation. Treating the Dirac equation in the limit of an
infinitely heavy quark $Q$ as an equation for a single light antiquark $\bar{q}$ (similarly for the case of hydrogen-like
atoms), we can study several important aspects of the theory of heavy-light quark-antiquark systems, in
particular, the relativistic dynamics of the light antiquark $\bar{q}$ in the external field of the heavy quark $Q$, the
Lorentz-structure of the long-range component of the $Q\bar{q}$ interaction, the fine structure of the spectrum of
heavy-light mesons, and the influence of the spontaneous breaking of chiral symmetry on the spectrum, etc.

The mathematical theory of the Dirac equation with a scalar-vector interaction was developed in \cite{Greiner}
(see \cite{Matsuki,Dvoeglazov,Eides} for a detailed bibliography). Certain progress was achieved in constructing the exact solutions of
equations of this type with potentials corresponding to different types of interaction \cite{Greiner}. However, in most cases attempts to construct exact solutions of this equation for more or less realistic potentials encounter difficulties that have not yet been overcome. The known methods of investigation of this equation approximately
(the perturbation theory in the coupling constant, etc.) do not provide complete knowledge about the behavior of the wave functions and mass spectrum in the most interesting domain of values of the coupling
constant for hadronic systems containing one light quark together with one ($D$ and $B$ mesons) or two (doubly heavy $\Xi$ and $\Omega$ baryons) heavy (anti)quarks; relativistic and nonperturbative effects evidently play an important role in such systems. Therefore, when constructing approximate methods for investigating bound
states of the Dirac equation, nonperturbative methods, in which the expansion parameter in the potential
is not considered small, are especially important. One of the most widely used methods is the
method of asymptotic expansion in the Planck constant $\hbar$, which is called the quasiclassical approximation.

The apparatus of quasiclassical asymptotic behavior which has arisen soon after making a quantum mechanics is now one of the most powerful and (frequently) also a unique way of study of a wide class of problems of theoretical and mathematical physics. Fundamental nature of a quasiclassical method is caused, in particular, by its embodiment of the important (for physical theories)  principle of correspondence of quantum and classical problems.

In the one-dimensional spectral problems of a quantum mechanics the strict mathematical formulation of this principle composes the Wentzel-Kramers-Brillouin approach (WKB). Historically first of all, namely the one-dimensional WKB approximation began to be applied for solving some of the quantum mechanical problems and in problems of the collisions theory. The WKB method has allowed to obtain general expressions for reflectivities and transition of particles through a potential barrier, for quasiclassical phases and scattering amplitudes in a central field, for Bohr-Sommerfeld quantization conditions, for probabilities of nonadiabatic transitions at pseudocrossing of levels.

The rigorous theory of quasiclassical asymptotic expansions including the scattering problem together
with spectral problems, was constructed in Maslov's fundamental monograph \cite{Maslov} and subsequent articles \cite{Maslov1}.
The WKB method for fermions satisfying the Dirac equation with a purely vector interaction (including
states lying near the boundary of the lower continuum) was developed in detail in \cite{Mur2,Mur3,Popov5}. No wonder that the framework of the quasiclassical methods has solely allowed to obtain many approved results in the known theory of superheavy atoms \cite{Popov_Rev}.

The construction of quasiclassical solutions of the spinor equation with a scalar-vector interaction was
reported in \cite{Simonov1,Lazur}. The scheme of  quasiclassical quantization proposed in \cite{Lazur} allows to make clear the connection of quasiclassical asymptotic behavior in spectral problems for the Dirac equation in external scalar and vector fields with the Lorentz structure of interaction potentials corresponding to them.

To elucidate how the Lorentz structure of interaction potentials is manifested at solving spectral problems of hadron physics, in the present article we study the behavior of a relativistic spin-1/2 particle in the presence of both the scalar and the vector external fields with potentials of confining type by WKB method. More definitely, at creating of the relativistic version of the potential model taking into account the Lorentz structure of potentials of interquark interaction, we relied on Cornell model offered in \cite{Eichten} in which the effective color Coulomb attraction at small distances $r$ and string interaction at large $r$ are considered. Corresponding to such model of $Q\bar{q} $-interaction the effective potential (EP) $U(r,E)$ of the Dirac equation complicatedly depends on energy $E$ and on mixing coefficient $\lambda$ ($0\leqslant\lambda\leqslant 1$) of scalar and vector long-range potentials and has essentially different form at $\lambda<1/2$ \cite{part1}, $\lambda>1/2$ \cite{Lazur_IJMPA} and $\lambda=1/2$. For all further consideration, the important moment here is that the same model with a scalar-vector variant of interactions leads (depending on range of values of the mixing parameter $\lambda$) both to bound states and to decaying (quasistationary) ones. Therefore, other field of application of the used potential model concerns to rather important problem related to study of quasistationary states of quantum objects. For this reason as an example of the chosen model of interaction, in the present article the problems of theory of decaying states are considered together with spectral problems.

The structure of the article is as follows. In the next section, we set out the construction of the quasiclassical solutions of the Dirac equation with a scalar-vector coupling. In Sec.~\ref{s3}, we consider the model of the interaction of a relativistic particle with both scalar and vector fields given by the potentials $V(r)=-\xi/r+\lambda(\sigma r+V_0)$) $S(r)=(1-\lambda)(\sigma
r+V_0)$ and study the dependence of the effective potential on the Lorentz structure of the external field. In Secs.~\ref{s4} and \ref{s5}, in the framework of the considered model with mixing parameter $0\leqslant\lambda<1/2$ we obtain simple asymptotic formulas for the energy and mass spectra and for the mean radii of $D$, $D_s$, $B$, and $B_s$ mesons. Our results are compared with results of numerical calculations and existing experimental data. Sec.~\ref{s6} is devoted to deriving asymptotic coefficients of the wave function at the zero and infinity in quasiclassical approximation. In Sec.~\ref{s11} we calculate the energy spectrum of the massless fermion in the external scalar field described by the combined funnel potential.

\section{quasiclassical approximation for the Dirac equation with a vector and scalar interaction potential}\label{s2}

After separation of angular variables in the Dirac equation with a mixed scalar-vector coupling ($c=1$), system of equation
for the radial wave functions $F(r)$ and $G(r)$ is of the form
\begin{equation}\label{1}
\left.\begin{array}{c}
\displaystyle\hbar\frac{dF}{dr}+\frac{\tilde{k}}{r}F-
\left[\left(E-V(r)\right)+\left(m+S(r)\right)\right]G=0,\vspace{1mm}\\
\displaystyle\hbar\frac{dG}{dr}-\frac{\tilde{k}}{r}G+\left[\left(E-V(r)\right)-
\left(m+S(r)\right)\right]F=0.\end{array}\right\}
\end{equation}
Hereafter, we use the notation $F(r)=rf(r)$ and
$G(r)=rg(r)$, where $f(r)$ and $g(r)$ are the radial
functions for the respective upper and lower components of the Dirac bispinor \cite{Ahiezer}, $E$ and $m$ are the total
energy and rest mass of the particle, $S(r)$ is the Lorentz-scalar potential, and the potential $V(r)$ up to a
multiplier coincides with the zeroth (temporal) component of the four-vector potential $A_{\mu}=(A_0,\bf A)$, where
${\bf A}=0$, $V(r)=-eA_0(r)$, and $e>0$. In system (\ref{1}), $\tilde{k}=\hbar\,k$, where the quantum number
$k=\mp(j+1/2)$ for $l=j\mp1/2$, $j$ is the total angular moment of the fermion, and $l$ is the orbital moment (for the upper component of $F(r)$), and hence $|k|=j+1/2=1,2,\ldots$.

\begin{figure}[b]
\vspace*{-1mm}
\centerline{\includegraphics[width=80mm]{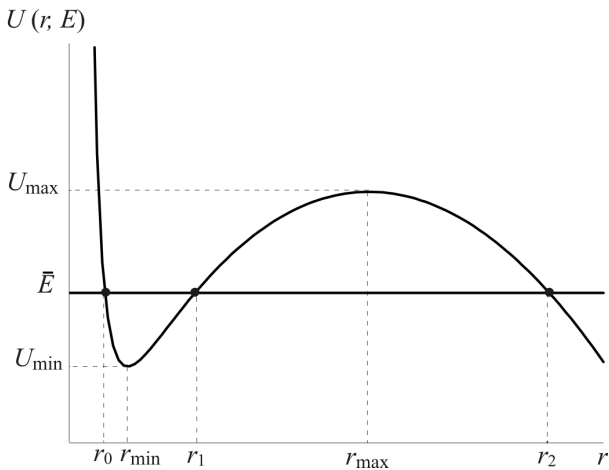}}
\caption{\footnotesize The form of the EP $U(r,E)$
of the barrier type; $r_0$, $r_1$, and $r_2$ are roots of the equation $p^2=0.$
}\label{f1}
\end{figure}

In \cite{Lazur} the system (\ref{1}) was consecutively solved using the known technique of left and right eigenvectors of
the homogeneous system. For the effective potential (EP) of the barrier type (see Fig. \ref{f1})
\begin{equation}\label{2}
U\left(r,E\right)=\frac{E}{m}V+S+\frac{S^{2}-V^{2}}{2m}+\frac{k^{2}}{2m\,r^{2}}
\end{equation}
quasiclassical expressions were obtained for the wave functions in three regions.

i) in the classically allowed region $r_0 < r < r_1$ the asymptotic approximation of the WKB radial
functions $F$ and $G$ are oscillatory:
\begin{equation}\label{2a}
\begin{array}{l}
F(r)=C_{1}^{\pm}\left[\displaystyle\frac{E-V+m+S}{p(r)}\right]^{1/2}\cos
\Theta_{1},\\
G(r)=C_{1}^{\pm}\mathrm{sgn}\,k\left[\displaystyle\frac{E-V-m-S}{p(r)}\right]^{1/2}\cos
\Theta_{2},
\end{array}
\end{equation}
where
\begin{equation}\label{2b}
p(r)=\left[(E-V(r))^{2}-(m+S(r))^{2}-(k/r)^2\right]^{1/2}
\end{equation}
is the quasiclassical momentum for the radial motion of a particle, and
\begin{eqnarray*}
\Theta_{1}(r)=\int\limits^{r}_{r_{1}}\left(p+\frac{k\,w}{p\,r}\right)dr+\frac{\pi}{4},\\
\Theta_{2}(r)=\int\limits^{r}_{r_{1}}\left(p+\frac{k\,\tilde{w}}{p\,r}\right)dr+\frac{\pi}{4},\\
w=\frac{1}{2}\left(\frac{V'-S'}{m+S+E-V}-\frac{1}{r}\right),\\
\tilde{w}=\frac{1}{2}\left(\frac{V'+S'}{m+S-E+V}+\frac{1}{r}\right).
\end{eqnarray*}
Hereafter, the normalization constants
$C_j$ related to the states with $k > 0$ and $k < 0$ are denoted by the superscripts $+$ and $-$. The normalization constant
$C_1^{\pm}$ is determined by the relation
\begin{equation}\label{t}
\left|C_{1}^{\pm}\right|=\left\{\int\limits^{r_{1}}_{r_{0}}\frac{E-V(r)}{p(r)}dr\right\}^{-1/2}=
\left(\frac{2}{T}\right)^{1/2},
\end{equation}
where the quantity $T$ coincides with the period of radial oscillations of a classical relativistic
particle with the energy $E$ in the potential well $r_{0}<r<r_{1}$.

ii) in the below-barrier region $r_1 < r < r_2$, the quantity $p$ takes purely imaginary values, $p = iq$,
$q=\sqrt{(m+S(r))^{2}-(E-V(r))^{2}+(k/r)^2}$.
Oscillating-type WKB solution (\ref{2a}) is then continued here
by a solution exponentially decreasing as the distance increases in the classically forbidden region $r_1 < r < r_2$. For states with $k > 0$, we have
\begin{align}
\chi=\displaystyle\frac{C^{+}_{2}}{\sqrt{qQ}}\exp\left\{-\int\limits^{r}_{r_{2}}
\left[q+\frac{(m+S)V'+(E-V)S'}{2\,qQ}\right]dr\right\}\nonumber\\
\times\left(
\begin{array}{c}
-Q \\ m+S-E+V\end{array}\right),&\label{2e}
\end{align}
where $Q=q+|k|r^{-1}$. The solution for the states with $k < 0$ can be found in \cite{Lazur}.

iii) in the ``exterior'' classically allowed region $r > r_2$, a quasistationary state is associated with the
divergent wave
\begin{align}\label{2k}
\chi=\frac{C^{+}_{3}}{\sqrt{pP}}\exp\left\{\int\limits^{r}_{r_{2}}
\left[ip+\frac{(m+S)V'+(E-V)S'}{2\,pP}\right]dr\right\}\nonumber\\
\times\left(
\begin{array}{c}
      iP \\ m+S-E+V \end{array} \right).
\end{align}
This expression must be used to study the states with $k > 0$ ($k<0$ see \cite{Lazur}); here, $P=p+i|k|r^{-1}$ and the radial
momentum $p(r)$ is again positive.

The quasiclassical representations (\ref{2a})--(\ref{2k}) constructed are invalid in small neighbourhoods of turning points $r_j$
($j=0,1,2$). To bypass these points and match the solutions one can use the Zwaan method \cite{Zwaan} that allows to establish relations between normalization constants:
\begin{align}\label{2m}
C_{2}^{\pm}&=-i\,C_{3}^{\pm}=\mp\displaystyle
\frac{C_{1}^{\pm}}{2}
\left[\frac{E-V\left(r_1\right)+m+S\left(r_1\right)}
{\left|k\right|r_1^{-1}}\right]^{\pm\frac{1}{2}} \nonumber \\
&\times\exp \displaystyle
\left\{-\int\limits_{r_{1}}^{r_{2}}\left[q \pm
\frac{\left(m+S\right)V'+\left(E-V\right)S'}{2\,qQ}\right]dr\right\}.
\end{align}

Also in \cite{Lazur} the following quantization condition, determining the energy (position) of the bound state $E$ in the
mixture of the scalar and vector potentials, was obtained:
\begin{equation}\label{3a}
\begin{array}{c}
\displaystyle
\int\limits_{r_{0}}^{r_{1}}\left(p+\frac{k\,w}{p\,r}\right)dr=
\left(n_{r}+\frac{1}{2}\right)\pi.
\end{array}
\end{equation}
Here, $n_r=0,1,2,\ldots$ is the radial quantum number. The new quantization rule (\ref{3a}) differs from the standard Bohr-Sommerfeld quantization condition \cite{Landau} by the relativistic expression for the momentum $p(r)$ and by the correction proportional to $w(r)$,
which takes
into account the spin-orbital interaction and results in the splitting of levels with different signs of the
quantum number $k$.

Having calculated the flux of the particles outgoing to infinity by means of quasiclassical formulas (\ref{2k}), (\ref{2m}), we find the following expression for the level width:
\begin{equation}
\Gamma=\frac{1}{T}\exp\left[-2\,\Omega\right].
\label{eq24a}
\end{equation}
\begin{eqnarray}
&T=\displaystyle 2\int\limits_c^b\frac{E_r-V}{p}\,dr,\quad
\Omega=\displaystyle \int\limits_b^a
\left(q-\frac{kw}{q\,r}\right)dr,& \label{eq25}
\end{eqnarray}
which is valid for both signs of $k$ \cite{Lazur}.

The obtained quasiclassical formula (\ref{eq24a}) is the relativistic generalization of the well-known Gamow formula for the width of a quasistationary level. The nontrivial moment of such a generalisation is the modification of expression for the period of oscillations $T$ and the occurrence of the additional factor in the preexponent of expression (\ref{eq24a}) that depends on a sign of the quantum number $k$ and is caused by the spin-orbit coupling in the mixture of the scalar $S(r)$ and vector $V(r)$ potentials.

The described scheme of quasiclassical quantization is applicable for deriving the asymptotic behaviors of level energy and corresponding eigenfunctions, generally speaking, at ``large'' quantum numbers. However, in well-known exactly solvable spectroscopic problems of the relativistic quantum mechanics, for example for scalar $S (r) $ and vector $V(r)$ potentials of Coulomb ($S(r)=-\xi'/r$, $V(r)=-\xi/r$) or oscillatory ($S(r)=V(r)=\omega\,r^2/4$, $\omega>0$) types, formulas of quasiclassical quantization reproduce an energy spectrum precisely (even for lowest states) \cite{Lazur}. This circumstance allows to suppose that quasiclassical calculation methods developed in present article will be useful also for problems of hadronic spectroscopy in the range of small quantum numbers.

The spectral problem for the Dirac equation with the potentials $S(r)$ and $V(r)$ of the confining type that considered in subsequent sections illustrates applying these methods to problems in hadronic physics. Other
types of the potentials $S(r)$ and $V(r)$ and also a more detailed mathematical description of the WKB method
for the Dirac equation with a scalar-vector interaction can be found in \cite{Lazur}.

\section{The dependence of the EP $\boldsymbol{U(r,E)}$ on the Lorentz structure of the external field}\label{s3}

The simplest model of the interaction of a relativistic spin-1/2 particle simultaneously with both scalar
and vector external fields, which we meet below when calculating the quasiclassical spectrum of relativistic
bound states (see Sec. \ref{s4}), is governed by the potentials
\begin{equation}\label{potential}
\begin{array}{l}
\displaystyle V(r)\equiv V_{\mathrm{Coul}}(r)+V_{\mathrm{l.r.}}(r)=-\frac{\xi}{r}+\lambda v(r), \\
\displaystyle S(r)\equiv S_{\mathrm{l.r.}}(r)=(1-\lambda) v(r), \quad v(r)=\sigma r+V_0,
\end{array}
\end{equation}
where $V_0$ is a real constant, $\xi$ is the Coulomb coefficient, and
$\lambda$ is the parameter of mixing between the
vector and scalar long-range potentials $V_{\mathrm{l.r.}}(r)$ and $S_{\mathrm{l.r.}}(r)$; $0\leqslant\lambda\leqslant1$. Below in this section, we do not
restrict the value or even the sign of the parameter $\sigma$.

The relation between the EP $U(r, E)$ and initial potentials (\ref{potential}) directly entering the Dirac equation is
rather complicated: $U(r, E)$ depends not only on $r$ and model parameters (\ref{potential}) but also on the level energy $E$
and on the total moment $j$. What is especially important for us here is that the EP $U(r, E)$ takes essentially
different forms for the cases $\lambda<1/2$,
$\lambda>1/2$, and $\lambda=1/2$.

Our goal is to investigate the behavior of the EP $U(r, E)$ at large and small $r$. Substituting $V(r)$ and $S(r)$ of form (\ref{potential}) in
(\ref{2}) and keeping only the most singular terms when $r\rightarrow 0$ only the leading terms (in $r$) when $r\rightarrow \infty$, we obtain:
$$ U\left(r,E\right)\sim\left\{\begin{array}{llr}
\displaystyle\frac{(1-2\lambda)\sigma^2}{2m}\,r^2+\ldots, \,
r\rightarrow\infty,& \lambda \neq \displaystyle\frac{1}{2},&\,(13\mbox{a})\vspace{1mm}\\
\displaystyle\frac{E+m}{2m}\,\sigma\,r+\ldots, \,
r\rightarrow\infty,& \lambda =
\displaystyle\frac{1}{2},&\,(13\mbox{b})
\vspace{1mm} \\\displaystyle\frac{\gamma^2}{2m\,r^2}, \, r\rightarrow0,
\, \gamma^2=k^2-\xi^2.&& \,(13\mbox{c})
\end{array}\right.\setcounter{equation}{13}
$$
For rather large values of the Coulomb coupling constant $\xi$ in EP $U(r,E)$ at small distances $r $ the singular attraction $\propto r^{-2}$ arises, which can lead to ``falling to center'' known in quantum mechanics \cite{Landau,Case,Perelomov}. In order to show this, we consider the finite trial function which is nonzero in the range $0 <r <r_0$. In accordance with the Heisenberg's indeterminacy relation $\left\langle p^2\right\rangle r^2_0\geqslant 1/4$, whence follows
\[\left\langle H \right\rangle=\frac{\left\langle
p^2\right\rangle}{2 m}+\left\langle U\right\rangle \leqslant
\frac{(j+1/2)^2-\xi^2}{2 m r^2}+O\left(\frac{1}{r_0}\right), \quad
r_0\ll 1.
\]
At $\xi>|k|=j+1/2$ the spectrum of the effective Hamiltonian $H $ is unbounded below, because $ \left\langle H \right\rangle \rightarrow-\infty $ at $r_0 \rightarrow 0$. In the classical mechanics such situation corresponds to a particle falling to a force centre. In the case of the Dirac equation the energy eigenvalues $E _ {n_r k} $ have the squared-root singularity at $ \xi \rightarrow |k | $ that leads to states with complex energy $E _ {n_r k} $ at the formal continuation of the one-particle solutions into the domain $ \xi> |k | = j+1/2$. It does not mean, however, that at $ \xi> |k | $ the Dirac equation has no solutions. For determination of energy levels in this case it is necessary to impose some boundary condition at $r \rightarrow 0$ (that is equivalent to determination of the self-conjugate expansion of the energy operator \cite{Baz}). From the physical point of view the formulation of a boundary condition at zero means the cut-off of the (color-) Coulomb potential $V_{\mathrm{Coul}}(r)$ at small distances, i.e. taking into account finite sizes of an extended source (heavy quark, nucleus):
\begin{equation}\label{obriz}
V(r)=\left\{\begin{array}{ll} \displaystyle-\frac{\xi}{r}+\lambda
v(r) &\mbox{for}\quad r>r_N,\vspace{1mm}\\
\displaystyle-\frac{\xi}{r_N}\,f\left(\frac{r}{r_N}\right)&
\mbox{for} \quad 0<r<r_N
\end{array}\right.
\end{equation}
(here $r_N$ is the cut-off radius, and $f(0)<\infty$). The form of the cut-off function $f(r/r_N)$ is determined by distribution of (color) electrical charge over volume of a nucleus (quark) (see \cite{Mur4,Zeldovich}).

Under the requirement $r_N\ll|\sigma|^{-1/2}$ which is satisfied for small values of the Coulomb parameter $ \xi <|k | $, level energy $E _ {n_r k} $ depends weakly on a concrete form of $V (r) $ at small $r $. Therefore, we carry out the passage to the limit ($r_N \rightarrow 0$) of the pointlike Coulomb potential in (\ref{obriz}). If $ \xi> |k |$ then the dependence of $E _ {n_r k} $ on the form of cut-off function $f (x) $ becomes rather strong which is characteristic for all problems with ``falling to center''.

Let us now consider behavior of EP $U(r,E)$ and wave functions at large distances $r$ in more detail.
First note that only the quadratic term $(S^2-V^2)/2m$ is essential in the asymptotic domain
in formula (\ref{2})  for $\lambda\neq 1/2$ and has the behavior
$(1-2\lambda)\sigma^2r^2/2m$ when $r\rightarrow\infty$. It is hence obvious
that for any sign of the parameter $\sigma$, the EP $U(r,E)$ of model (\ref{potential}) under consideration (at sufficiently large
distances) is an attractive potential for $\lambda>1/2$ and a repulsive potential for $\lambda<1/2$. Both types of
behavior (i.e., attraction for $\lambda>1/2$ and repulsion for $\lambda<1/2$) are purely relativistic effects related to the fact that the interaction of the fermion with the scalar external field $S(r)$ is added to the scalar quantity $m$, the particle mass, while the vector potential $V(r)$ is introduced into the free Dirac equation minimally
as the temporal component of the Lorentz-vector $A_{\mu}$.

\begin{figure}[b]
\vspace*{-1mm}
\centerline{\includegraphics[width=65mm]{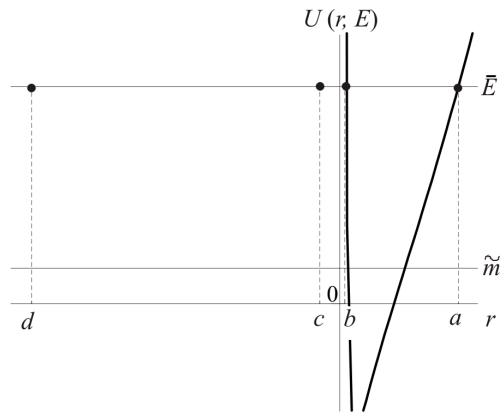}}
\caption{The EP $U(r,E)$ of Dirac system (\ref{1}) with potential (\ref{potential}) in the case where $\lambda<1/2$,
$\sigma>0$, and $\widetilde{E}>\widetilde{m}$; $a$, $b$, $c$ and $d$ are the quasimomentum roots in (\ref{A2}).
}\label{f1a}
\end{figure}

It is clear from what was said above that for $\lambda<1/2$, the EP $U(r,E)$ of model (\ref{potential}) is an unboundedly increasing (as $r$ increases) confining potential with only a discrete spectrum of energy levels; it is then essential that the quadratic dependence of the EP $U(r,E)$ on $r$ (and hence the confinement property)
appears because of the relativistic terms $(S^2-V^2)/2m$. An example form of the EP $U(r,E)$ for $\lambda<1/2$
is shown in Fig.~\ref{f1a}. It is amazing that bound states are present in composite field (\ref{potential}) under consideration for $\lambda<1/2$ even in the case where the initial long-range potential $v(r)=\sigma r+V_0$ corresponds to attraction ($\sigma<0$, $V_0<0$).

But for $\lambda>1/2$ and an arbitrary value of $\sigma\neq 0$, the effective Hamiltonian $H$ of the squared Dirac
equation in external field (\ref{potential}) has complex eigenvalues of energy because the EP $U(r,E)$ becomes negative in this case (at sufficiently large distances) and less than the effective particle energy $\bar{E}=(E^2-m^2)/2m$,
which corresponds to attraction. Therefore, for $\lambda>1/2$, the EP $U(r,E)$ of model (\ref{potential}) has the form of a well separated from the external domain by a wide potential barrier (for $|\sigma|\ll 1$; see Fig.~\ref{f1}). It is obvious that the leading contribution to forming the barrier of the EP $U(r,E)$ comes from the Lorentz-vector
component $V_{\mathrm{l.r.}}(r)$ of the long-range potential $v(r)$. Furthermore, as follows from (\ref{2}) and (15a), in the presence of only a vector field ($\lambda=1$), the EP $U(r,E)$ does not have the confining property even when the initial long-range potential $v(r)=\sigma r+V_0$ corresponds either to attraction ($\sigma<0$, $V_0<0$) or to repulsion ($\sigma>0$, $V_0>0$). This is the principal difference between relativistic potential model (\ref{potential}) under consideration and the analogous nonrelativistic model in which the EP $U^{n.r.}_{eff}(r)=-\xi/r+\sigma r+V_0+l(l+1)/2r^2$ in the radial Schr\"{o}dinger equation has the barrier for negative values of the parameters $\sigma$ and $V_0$, which results in quasistationary states with complex energies appearing instead of discrete levels. On the contrary, if $\sigma>0$, then the EP $U^{n.r.}_{eff}(r)$ becomes an unboundedly increasing confining potential with only the discrete spectrum of energy levels. The absence of bound states in the Dirac equation with a linearly increasing vector potential $V(r)$ was first noted in \cite{Plesset}.

The quasiclassical formulas for the wave functions in the domain $r>r_2$ \cite{Lazur} imply the asymptotic
form of the radial functions $F(r)$ and $G(r)$ as $r\rightarrow\infty$. It then happens that the wave functions decrease exponentially at large distances for $0\leqslant\lambda<1/2$ and oscillate if $1/2<\lambda \leqslant1$. As an illustration, we present this asymptotic behavior for the radial function corresponding to the upper component of the Dirac bispinor ($r\rightarrow\infty$):
\begin{equation}\label{asympt1}
F\sim\left\{\begin{array}{ll} \displaystyle
\exp\left(-\frac{\sqrt{1-2\lambda}\,|\sigma|}{2}\,r^2\right)\,
& \mbox{for}\,\, 0\leqslant\lambda<1/2,\\
\displaystyle\exp\left(i\frac{\sqrt{2\lambda-1}\,|\sigma|}{2}\,r^2\right)\,
& \mbox{for}\,\, 1/2<\lambda\leqslant 1.
\end{array}\right.
\end{equation}
It hence follows that the relativistic solutions for potential model (\ref{potential}) (depending on the value of the mixing parameter $\lambda$) constitute stationary or quasistationary systems satisfying different boundary conditions (\ref{asympt1}) for $\lambda < 1/2$ or $\lambda > 1/2$.

We point out one more important particular case realized at $\lambda=1/2$. Substituting potentials (\ref{potential}) with the value $\lambda=1/2$ in expression (\ref{2}), we see that the quadratic dependence of the ``tail'' of $U(r,E)$ on $r$ disappears and the long-range components $V_{\mathrm{l.r.}}(r)$ and $S_{\mathrm{l.r.}}(r)$ of the first two terms dominate EP (\ref{2}) at large $r$, which results in a practically linear dependence of $U(r,E)$ on $r$ (see (15b)). We note that we again obtain a linear confining potential, which has only the discrete spectrum, at positive values of $\sigma$, while for negative (sufficiently small) values of $\sigma$, the EP $U(r,E)$ of model (\ref{potential}) has a wide barrier. Because of this, level decay by percolation through the potential barrier becomes possible, i.e., the bound level becomes a quasistationary exponentially decaying state with the complex energy $E=E_{r}-i\Gamma/2$. From the analyticity standpoint, the above behavior of the EP $U(r,E)$ for $\sigma<0$ and $\sigma>0$ allows studying how the discrete spectrum continues from the real axis to the complex plane.

Summarizing, we can say that varying one of the parameters of interaction model (\ref{potential}), the coefficient
$\lambda$ of mixing the scalar and vector long-range potentials $S_{\mathrm{l.r.}}(r)$ and $V_{\mathrm{l.r.}}(r)$, in the interval $0\leqslant\lambda\leqslant1$, we obtain qualitatively different forms of the EP $U(r,E)$: from the confining potential with only the discrete spectrum for $\lambda < 1/2$ to the potential with the potential barrier and quasistationary energy levels for $\lambda < 1/2$ through the physically important intermediate case $\lambda = 1/2$, where the asymptotic behavior (as $r\rightarrow \infty$) of the ``tail'' of the EP $U(r,E)$ switches from quadratic (15a) to linear (15b) (see above).

For the squared Dirac equation in composite field (\ref{potential}), the form of the EP becomes more complicated:
expression (\ref{2}) for $U(r,E)$ acquires small corrections due to the particle spin and the related spin-orbital
interaction. It is clear from the nature of the conclusions about the behavior of the EP $U(r,E)$ for $\lambda < 1/2$,  $\lambda > 1/2$, and $\lambda = 1/2$ that the indicated changes of the form of $U(r,E)$ do not change the results
qualitatively.

Everything said above remains valid for the spherically symmetric potentials $S(r)$ and $V (r)$ with the
powerlike or logarithmic behavior ($v(r)\sim \sigma r^{\beta}$, $\beta>0$, or $v(r)\sim g\log r$) of the long-range part $v(r)$ at infinity.

Having clarified the qualitative aspects, we now concentrate on a practical application of the above
apparatus of quasiclassical asymptotic behavior to heavy-light mesons.

\section{quasiclassical description of the energy spectrum of heavy-light quark-antiquark systems}\label{s4}

To use the potential approach to describe properties of heavy-light mesons, we must construct the
quark-antiquark interaction potential. As is known from QCD, because of the asymptotic freedom property,
the Coulomb-type potential of the one-gluon exchange gives the leading contribution at small distances
($r<0.25$Fm).

As the distance increases, the long-range confining interaction (the confinement), whose actual form has
not yet been established in the QCD framework, prevails. The confining potential may have a complicated
Lorentz structure. For example, it was shown in \cite{Dosch,Simonov} that the interaction of the quark-antiquark pair
with a fluctuating gluon vacuum field at a finite correlation length results in a linearly increasing potential.
The spin-dependent potential obtained with that approach has a structure that is characteristic of scalar
confinement. On the other hand, the infrared asymptotic behavior of the gluon propagator of the form
$D({\bf k}^2)\sim 1/({\bf k}^2)^2$ was obtained in \cite{Arbuzov} by analyzing the system of the Schwinger-Dyson equations. Such an asymptotic behavior in the static limit results in a linearly growing vector confining potential. It is therefore most plausible that the confining potential comprises a mixture of vector and scalar parts. Moreover, lattice calculations \cite{Otto} based on the first principles of QCD support a linear confinement proportional to $r/4\pi\alpha'(0)$ (where $\alpha'(0)$ is the slope of the hadronic Regge trajectory). From the above considerations, we assume that the $Q\bar{q}$ interaction is a combination of the following potentials:

\textbf{a}. the one-gluon exchange potential $V_{\mathrm{Coul}}(r)=-\xi/r$, where $\xi=4/3\,\alpha_s$, $\alpha_s$ is the strong coupling constant
\begin{equation}\label{asympfr}
\alpha_s(Q)=12\pi/[(33-2N_f)\log(Q^2/\Lambda^2)],
\end{equation}
$N_f$ is the number of quark flavors, and $\Lambda=360$\,MeV is the QCD parameter,

\textbf{b}. the long-range linear scalar confining potential $S_{\mathrm{conf}}(r)$ =$(1-\lambda)\,v(r)$, where $v(r)$ is determined by expression (\ref{potential}), and

\textbf{c}. the long-range linear vector potential $V_{\mathrm{conf}}(r)=\lambda\,v(r)$.

The total effective quark-antiquark interaction is then described by a combination of the perturbative one-gluon
exchange potential $V_{\mathrm{Coul}}(r)$  and the scalar and vector long-range confining potentials $S_{\mathrm{conf}}(r)$ and $V_{\mathrm{conf}}(r)$. Therefore, the potentials $S$ and $V$ are given by (\ref{potential}), where $\sigma=0.18\,\mbox{GeV}^2$ is the string tension, $V_0$ is the constant of the additive shift of the bond energy, and the coefficient $\lambda$ of mixing between the vector and scalar confining potentials is the adjustable parameter, $0\leqslant
\lambda<1/2$. We can consider that the value of $\alpha_s$ is approximately the same for each family of heavy-light mesons and doubly heavy barions and changes in accordance with (\ref{asympfr}) only when we pass from one family to another.

In the nonrelativistic limit ($E\approx m$ and $S,|V|\ll m$) EP becomes
\begin{equation}\label{U_nonrel}
U_{eff}\approx S(r)+V(r)+\frac{(l+1/2)^2}{2 m r^2}
\end{equation}
where the sum
\begin{equation}\label{Kornel}
U_{n.r.}=S(r)+V(r)=-\frac{\xi}{r}+\sigma r+V_0
\end{equation}
plays the role of an interaction potential

Therefore, the description of $Q\bar {q}$-interaction by means of the representation (\ref {potential}) for potentials $V (r) $ and $S (r) $ is nothing else than a generalization of the interquark interaction potential (\ref{Kornel}) to the relativistic case of QCD-motivated Cornell model \cite{Eichten}.

As is known, the potential (\ref{Kornel}) possesses solely a discrete spectrum at $ \sigma> 0$ and is the spherical model of the Stark effect in the hydrogen atom at $\sigma <0$. In the nonrelativistic problem it is indifferent what of terms ($V(r)$ or $S(r)$) in (\ref{Kornel}) ensures the confinement of quarks. However, the analysis of the Dirac system (\ref{1}) with potentials (\ref{potential}) shows (see Sec.~\ref{s3}) that the radial wave functions $F(r)$ and $G(r)$ exponentially decreasing (at $r\rightarrow\infty $) and normalised properly can be gained only when $S_{conf}(r)>V_{conf}(r)$ at $0<r<\infty$.

We cannot solve Dirac system (\ref{1}) with potentials (\ref{potential}) exactly; we hence use the quasiclassical approximation method, which provides a high accuracy even for low-lying quantum numbers in the case of scalar
and vector fields of the Coulomb and oscillatory types \cite{Lazur}.

Choosing the mixing coefficient in the range $0\leqslant\lambda<1/2$ corresponds to the scalar confinement
prevailing. In this case, the EP $U(r,E)$ of our model has the form of a standard oscillator well with
a single minimum (at the point $r_{min}\approx\gamma^2/\widetilde{E}\xi$) and no maximums (see Fig.~\ref{f1a}). The equation $p^2=2m(\bar{E}-U(r,E))=0$ determining the turning points then results in the complete fourth-degree algebraic equation $r^{4}+f\,r^{3}+g\,r^{2}+h\,r+l=0$ with the coefficients
\begin{equation}\label{coeff}
\begin{array}{c}
f=\displaystyle\frac{2[\widetilde{m}\left(1-\lambda\right)+\widetilde{E}\lambda]}
{\left(1-2\lambda\right)\sigma},\,
g=-\frac{\widetilde{E}^{2}-\widetilde{m}^{2}-2\xi\sigma\lambda}
{\left(1-2\lambda\right)\sigma^{2}},\\
h=-\displaystyle\frac{2\widetilde{E}\xi}{\left(1-2\lambda\right)\sigma^{2}},\,
l=\displaystyle\frac{\gamma^{2}}{\left(1-2\lambda\right)\sigma^{2}}
\end{array}
\end{equation}
where $\widetilde{E}=E-\lambda V_0$, $\widetilde{m}=m+(1-\lambda)V_0$ are the characteristic parameters with the respective meanings of the ``shifted'' energy and the ``shifted''
mass. This equation has four real roots $d<c<b<a$ determined by the equalities
\begin{equation} \label{A2}
\begin{array}{c}
 a=-\displaystyle\frac{f}{4}+\frac{1}{2}\left(\Xi + \Delta_{+}
\right),\, b=-\displaystyle\frac{f}{4}+\frac{1}{2}\left(\Xi - \Delta_{+}
\right), \\\\
c=-\displaystyle\frac{f}{4}-\frac{1}{2}\left(\Xi-
\Delta_{-}\right), \, d=-\displaystyle\frac{f}{4}-\frac{1}{2}\left(\Xi +
\Delta_{-}\right).
\end{array}
\end{equation}
Here, we use the notation
\begin{eqnarray*}
&\Xi=\left[\displaystyle\frac{f^2}{4}-\frac{2g}{3}+\frac{u}{3}\left(
\frac{2}{Z}\right)^{1/3}+\frac{1}{3}\left(\frac{Z}{2}\right)^{1/3}
\right]^{1/2},&\\
&\Delta_{\pm}=\sqrt {F\pm
\displaystyle\frac{D}{4\Xi}}, \quad
Z=v+\sqrt{-4u^3+v^2},&\\
&F=\displaystyle\frac{f^2}{2}-\frac{4g}{3}-\frac{u}{3}\left(\frac{2}{Z}\right)^{1/3}
-\frac{1}{3}\left(\frac{Z}{2}\right)^{1/3},& \\
&D=-f^3+4fg-8h,\quad u=g^2-3fh+12l,& \\
&v=2g^3-9fgh+27h^2+27f^2l-72gl.
\end{eqnarray*}

For the potentials under consideration, the quasiclassical momentum is determined by equalities (\ref{2b}) and (\ref{potential}). Using formulas (\ref{A2}), we represent it in the form convenient for what follows ($\sigma>0$ and $\sigma<0$)
\begin{eqnarray}
 p(r)&=&\displaystyle|\sigma|\sqrt{1-2\lambda}\,\frac{R(r)}{r}\nonumber\\
 &=&\displaystyle|\sigma|\sqrt{1-2\lambda}
\frac{\sqrt{(a-r)(r-b)(r-c)(r-d)}}{r}.\label{eq4a}
\end{eqnarray}
We integrate in quantization condition (\ref{3a}) over the classically allowed domain between the two positive
turning points $r_{0}=b<r_{1}=a$, while the other two turning points ($d<c<0$) are in the nonphysical
domain $r<0$. Using formula (\ref{eq4a}), we transform quantization integrals (\ref{3a}) into the sum of the integrals
\begin{equation}\label{A4}
\begin{array}{l}
\displaystyle J_{1}=\int\limits_{b}^{a}p(r)dr\\
\displaystyle=-|\sigma|\sqrt{1-2\lambda}
\int\limits_{b}^{a}\frac{\left(r^{3}+fr^{2}+gr+h+lr^{-1}\right)}{R}dr,\\
\displaystyle J_{2}=\int\limits_{b}^{a}\frac{k\,w}{p(r)r}dr\\
\displaystyle=\frac{-k}{2|\sigma|\sqrt{1-2\lambda}}\left[\int\limits_{b}^{a}\frac{dr}{\left(r-\lambda_{+}\right)R}+
\int\limits_{b}^{a}\frac{dr}{\left(r-\lambda_{-}\right)R}\right],
\end{array}
\end{equation}
where we introduce the notation
\[\lambda_{\pm}=-\frac{\widetilde{E}+\widetilde{m}\mp
\sqrt{(\widetilde{E}+\widetilde{m})^2-4\sigma
\xi(1-2\lambda)}}{2\sigma\left(1-2\lambda\right)}.\]

Writing condition (\ref{3a}) in terms of $J_1$ and $J_2$ is advantageous compared with the initial representation because the integrals contained in $J_1$ and $J_2$ can be expressed in terms of complete elliptic integrals.

The particle energy spectrum is determined by quantization condition (\ref{3a}), which, after quantization
integrals (\ref{A4}) are evaluated (see Appendix), becomes the transcendental equation
\begin{eqnarray}
&&\displaystyle-\frac{2\sqrt{1-2\lambda}}{\sqrt{\left(a-c\right)\left(b-d\right)}}\left\{\frac{|\sigma|\left(b-c\right)^{2}}
{\Re}\left[N_{1}F\left(\chi\right)+N_{2}E\left(\chi\right)\right.\right.\nonumber\\
&&\displaystyle+ N_{3}
\Pi\left(\nu,\chi\right)+N_{4}\Pi\left(\frac{c}{b}\nu,\chi\right)\Biggr]+\displaystyle\frac{k}{2\left(1-2\lambda\right)|\sigma|}\nonumber\\
&&\times\left.\left[(b-c)\left(N_{5}
\Pi\left(\nu_{+},\chi\right)+N_{6}
\Pi\left(\nu_{-},\chi\right)\right)+N_{7}F\left(\chi\right)\right]\right\}\nonumber\\
&&
=\left(n_r+\displaystyle\frac{1}{2}\right)\pi, \label{5}
\end{eqnarray}
where $F(\chi)$, $E(\chi)$, and $\Pi(\nu,\chi)$ are the complete elliptic integrals of the respective first, second, and third kind (see formulas (A.1)). The mathematical details of calculating integrals of type (\ref{A4}) can be found
in \cite{Beitmen,Prudnikov}, and the expressions for $\nu$, $\chi$, $\nu_{\pm}$, $\Re$, and $N_i$ ($i=1,...,7$) are collected in Appendix because they are rather cumbersome.

Finding an ``exact'' solution of Eq. (\ref{5}) in the general case is, of course, impossible, but the situation
is simplified with the increase in the energy $E$ or in the approximation of ``weak'' long-range field (as
compared with the Coulomb field). The first case corresponds to the fact that for not too large (i.e., for
``intermediate'') values of the parameters î and ó (namely, for $\sigma\lesssim 0.2\,\mbox{GeV}^2$ and $0.3<\xi<0.8$), the condition $\widetilde{E}^2\gg \sigma\gamma$ is well satisfied for all possible values $E_{n_r k}$ of the heavy-light meson energy levels, and the second case is realized when the condition $\sigma\ll\xi \widetilde{m}^2$ is satisfied. In the framework of our consideration (i.e., for the physics of heavy-light mesons), only the first case is interesting, while the second case is most often encountered in approximate calculations of those properties of low-lying hadronic states that do not depend directly on the presence or absence of confinement.

A simple and often effective method for deriving asymptotic expansions of integrals of form (\ref{A4}) is
to expand a quasimomentum $p(r)$ in a small parameter, the interaction, and integrate the obtained series
term by term. We then indicate two special features of this procedure for calculating the integrals $J_1$ and
$J_2$ containing the small parameter. First, it is obvious from analyzing expressions (\ref{A2}) that in addition to
the level $E=m$, we must introduce one more characteristic energy level $\widetilde{E}=\widetilde{m}$, which divides the domains of applicability of the asymptotic approximations for the quantization integrals $J_1$ and $J_2$ obtained below. Using the relations $\widetilde{E}>\widetilde{m}$ and $\widetilde{E}<\widetilde{m}$, we can show that in these two domains of the spectrum, the motion is quasiclassical if the condition $\sigma\gamma/\widetilde{E}^2\ll 1$ is satisfied for $\widetilde{E}<\widetilde{m}$ and the condition $\sigma\gamma/\widetilde{E}^2\ll 1$ is satisfied
for $\widetilde{E}>\widetilde{m}$. This gives the possibility of obtaining expressions for $J_1$ and $J_2$ in elementary functions using the formal expansion of the quasimomentum in a power series in a small dimensionless parameter (which is $\sigma\gamma/\widetilde{E}^2\ll 1$ or $\sigma/\xi \widetilde{m}^2\ll 1$). Second, the further analysis depends essentially on the mutual positions of the turning points $a$, $b$, $c$, and $d$. Then, depending on the relative values of $\widetilde{E}$ and the level $\widetilde{m}$, we consider several typical situations.

\textbf{Case A}: Let $\sigma>0$ and the conditions $\sigma\ll \xi\widetilde{m}^2$ and $\widetilde{E}<\widetilde{m}$ be satisfied. This situation describes deep levels whose energy is close to the bottom of the scalar-vector well $U(r,E)$. Estimating expressions (\ref{A2}) for the turning points in the approximation $\sigma/\xi \widetilde{m}^2\ll 1$ and preserving only the two first terms in the small parameter expansion, we can easily obtain
\begin{eqnarray}
&&\displaystyle a\approx\frac{\widetilde{E}\xi+ \theta}
{\mu^2}\left[1-\frac{\widetilde{E}\xi+
\theta}{\mu^4}\left(\eta_1+\frac{\widetilde{m}\xi\eta_2}{\mu}\right)\sigma\right],\nonumber\\
&&\displaystyle b\approx\frac{\widetilde{E}\xi-\theta}
{\mu^2}\left[1-\frac{\widetilde{E}\xi-
\theta}{\mu^4}\left(\eta_1-\frac{\widetilde{m}\xi\eta_2}{\mu}\right)\sigma\right],\label{root1}\\
&&\displaystyle c\approx-\displaystyle\frac{\widetilde{m}-\widetilde{E}}{\sigma}-\frac{\xi}{\widetilde{m}-\widetilde{E}},\,
d\approx\displaystyle-\frac{\widetilde{m}+\widetilde{E}}{\sigma(1-2\lambda)}+\frac{\xi}
{\widetilde{m}+\widetilde{E}}.\nonumber
\end{eqnarray}
Hereafter, we use the notation
\begin{equation}\label{theta}
\begin{array}{c}
\theta=\sqrt{(\widetilde{E}\,k)^2-(\widetilde{m}\,\gamma)^2},\quad
\mu=\sqrt{\widetilde{m}^2-\widetilde{E}^2}, \\
\eta_1=(1-\lambda)\widetilde{m}+\lambda\widetilde{E},\quad
\eta_2=\lambda \widetilde{m}+(1-\lambda)\widetilde{E}.
\end{array}
\end{equation}

It follows from asymptotic expressions (\ref{root1}) that the positive turning points $a$ and $b$ depend weakly
on $\sigma$ and are determined only by the Coulomb field. The other two (negative) turning points $c$ and $d$
depend mainly on the linear part $v(r)$ of interaction (\ref{potential}), but their values are ``corrected'' by the quantities $\mp\xi/(\widetilde{m}\mp\widetilde{E})$, which are due to the Coulomb long-range interaction. It is also obvious from (\ref{root1}) that for small positive values of $\sigma$, the turning points $c$ and $d$ are sufficiently far from the two points $a$ and $b$ and tend to $-\infty$ in the limit as $\sigma\rightarrow 0$.

The properties of deeply lying levels for massive quarks ($\widetilde{m}^2\xi\gg\sigma$) are mainly determined by the
Coulomb potential. Treating the long-range potential $v(r)$ as a small perturbation, we can expand the
quasiclassical momentum $p(r)$ in the domain of the potential well $b < r < a$ in a series in increasing powers
of $r/|c|\ll 1$ and $r/|d|\ll 1$. Calculating the table integrals in (\ref{3a}), whose sum gives the value of the
quantization integrals $J_1$ and $J_2$ up to terms of the order $O((\sigma/\xi\widetilde{m}^2)^2)$, we then obtain the equation, which can be easily solved for the level energies,
\begin{eqnarray}
\displaystyle E_{n_r k}&=&\widetilde{E}_0+\lambda
V_0+\frac{\sigma}{2\xi\widetilde{m}^2}
\left[\left(\frac{\xi^2\widetilde{m}^2}{\mu_0^2}-k^2\right)
\eta_{10}\right.\nonumber\\
&+&\displaystyle\left.\left(\frac{2\xi^2\widetilde{m}
\widetilde{E}_0}{\mu_0^2}-k\right)
\eta_{20}\right]+O\left(\left(\frac{\sigma}{\xi\widetilde{m}^2}\right)^2\right),\label{eq15a}
\end{eqnarray}
where $\widetilde{E}_0=\widetilde{m}\left[1+\xi^2/\left(n_r'+\gamma\right)^2\right]^{-1/2}$ is the Dirac level of the energy of the fermion (with the effective mass $\widetilde{m}=m+(1-\lambda)V_0$) in the Coulomb field, $n_r'=n_r+(1+\mathrm{sgn}\,k)/2$, and the quantities $\mu_{0}$, $\eta_{10}$, and $\eta_{20}$ are obtained from $\mu$, $\eta_1$, and $\eta_2$ by substituting $\widetilde{E}$ for $\widetilde{E}_0$. The previously accepted condition
$\sigma>0$ is unnecessary here because this result remains applicable also in the case of negative values of the
parameter $\sigma$.

Formula (\ref{eq15a}) can also be found using the standard perturbation theory, but this involves rather cumbersome
calculations. Using quasiclassical formulas (\ref{3a}) and (\ref{A4}) dramatically simplifies calculations. As
is shown by comparing with the result obtained by numerically integrating Eq. (\ref{1}), formula (\ref{eq15a}) ensures
a good accuracy for calculating the spectra of bound systems of heavy quarks (for example, $Q\bar{Q}$ mesons;
see \cite{Anikin}).

Calculations, which we omit here, demonstrate that in the case $\lambda<1/2$ and for (sufficiently small)
negative values of $\sigma$, the EP $U(r,E)$ has the shape of a double well. If we neglect the barrier penetrability
in the region $c<r<b$ between the two wells, then the quasiclassical quantization conditions in this well
can be written merely as the conditions on the phase integrals over the domain of the quasiclassical motion
in each of the wells. Quantizing in the left well by formula (\ref{3a}) then results in formula (\ref{eq15a}) above.

\textbf{Case B}: In the domain $\widetilde{E}>\widetilde{m}$ and $\sigma>0$, which is of actual importance for the physics of heavy-light mesons, a small dimensionless parameter $\sigma\gamma/\widetilde{E}^2$ appears in the spectral problem. Imposing the condition $\sigma\gamma/\widetilde{E}^2\ll 1$, we can easily obtain the approximate expressions for the turning points from exact formulas (\ref{A2}):
\begin{equation}\label{eq6}
\begin{array}{c}
a\approx\displaystyle\frac{\widetilde{E}-\widetilde{m}}{\sigma}+\frac{\xi}{\widetilde{E}-\widetilde{m}},
\qquad b\approx\frac{-\widetilde{E}\xi+ \theta}{\widetilde{E}^2-\widetilde{m}^2},\\
\displaystyle c\approx\frac{-\widetilde{E}\xi- \theta}{\widetilde{E}^2-\widetilde{m}^2},\qquad
d\approx-\frac{\widetilde{E}+\widetilde{m}}{\sigma(1-2\lambda)}+\frac{\xi}
{\widetilde{E}+\widetilde{m}}.
\end{array}
\end{equation}
As can be seen from these formulas, the turning points $a$ and $b$ are rather distant from each other, and the
above expansion for the quasimomentum $p(r)$ is not applicable in the whole integration domain. Nevertheless,
using the condition $\sigma\gamma/\widetilde{E}^2\ll 1$, we can use the approximation method to evaluate the quantization integrals based on the idea of splitting the whole integration domain [$b, a$] into the intervals [$b$, $\widetilde{r}$] and [$\widetilde{r}$, $a$] in each of which only the dominating interaction type is taken into account exactly while the other integration types are treated as perturbations.

We now find a point $\widetilde{r}$ that divides the integration domain $b\leqslant r\leqslant a$ into the domain $b\leqslant r \leqslant\widetilde{r}$ where the Coulomb potential prevails and the domain $\widetilde{r}\leqslant r \leqslant a$ where the long-range potential $v(r)$ prevails. The method for choosing such a point is not unique. The most natural seems to find a point $\widetilde{r}$ where the long-range potential $v(r)$ is equal to the Coulomb potential. From this requirement, we have $\widetilde{r}\approx(\widetilde{E}\xi/\eta_1\sigma)^{1/2}$.

We can calculate the quantization integrals (for $\sigma\gamma/\widetilde{E}^2\ll 1$) as follows. We calculate integrals (\ref{A4}) by expanding the quasimomentum $p(r)$ in a power series in the parameters $r/a\ll 1$ and $r/|d|\ll 1$ in the domain $b\leqslant r \leqslant\widetilde{r}$ and in the small parameters $b/r\ll 1$ and $|c|/r\ll 1$ in the domain $\widetilde{r}\leqslant r \leqslant a$. Splitting the integration interval at the point $\widetilde{r}\approx(\widetilde{E}\xi/\eta_1 \sigma)^{1/2}$ therefore gives the representation for $J_1$,
\begin{equation} \label{eq7}
J_{1}=\sigma\sqrt{1-2\lambda}\left(j_1+j_2\right),
\end{equation}
where the integrals $j_1$ and $j_2$ can be written as follows up to terms of the first order in the corresponding
small parameters $r/a$, $r/|d|$  and $b/r$, $|c|/r$ in the expansions for the quasimomentum $p(r)$:
\begin{equation}\label{eq8}
\begin{array}{l}
\displaystyle j_1=\sqrt{-a d}\int\limits_{b}^{\widetilde{r}}\frac{\sqrt{(r-b)(r-c)}}{r}\left[1-\frac{a+d}{2a
d}r+...\right]dr, \\
\displaystyle j_2=\int\limits_{\widetilde{r}}^{a}\sqrt{(a-r)(r-d)}\left[1-\frac{b+c}{2\,r}+\ldots\right]dr.
\end{array}
\end{equation}
Calculating the table integrals in (\ref{eq8}) and collecting terms with like dependence on $\sigma$, we obtain
\begin{eqnarray}
\displaystyle J_1&=&\sigma\sqrt{-ad(1-2\lambda)}
\left[\frac{b+c}{2}\log\left(\frac{(a-d)(c-b)}{16\,a\,
d}\right)\right.\nonumber\\
\displaystyle&-&\sqrt{-bc\,}\arccos\left(\frac{b+c}{b-c}\right)+\frac{a+d}{4}\nonumber\\
\displaystyle&+&\frac{1}{4\sqrt{-ad}}\left(\frac{(a-d)^2}{2}-(a+d)(b+c)\right)\nonumber\\
\displaystyle&\times&\left.\arccos\left(\frac{d+a}{d-a}\right)\right]+O\left(\frac{\sigma\gamma}{\widetilde{E}^2}\right).\label{eq10}
\end{eqnarray}
We note that when the asymptotic expressions for $j_1$ and $j_2$ are added, the result does not contain the
parameter $\widetilde{r}$.

To expand the integral $J_2$ in the small parameter $\sigma\gamma/\widetilde{E}^2$, we represent it as a sum of two terms,
\begin{equation} \label{eq11}
J_{2}=-\frac{k}{2\sigma\sqrt{1-2\lambda}}\left(\widetilde{j_1}+\widetilde{j_2}\right),
\end{equation}
where the integrals $\widetilde{j_1}$ and $\widetilde{j_2}$ can be written in the forms
\begin{equation}\label{eq12}
\begin{array}{l}
\displaystyle \widetilde{j_1}\simeq \frac{1}{\sqrt{-a\,d\,}}
\int\limits_{b}^{\widetilde{r}}\frac{dr}
{(r+\widetilde{p})\sqrt{(r-b)(r-c)}}, \\
\displaystyle \widetilde{j_2}\simeq
\int\limits_{\widetilde{r}}^{a}\frac{1}{\sqrt{(a-r)(r-d)}}\left[\frac{1}{r^2}+\frac{1}
{r(r+\widetilde{q})}\right]dr,
\end{array}
\end{equation}
where $\widetilde{p}=\xi/(\widetilde{E}+\widetilde{m})$, $\widetilde{q}=(\widetilde{E}+\widetilde{m})/\sigma(1-2\lambda)$.
An elementary calculation of the integrals in (\ref{eq12}) results in
\begin{eqnarray}
\displaystyle J_{2}&=&-\frac{k}{2|\sigma|\sqrt{1-2\lambda}}\frac{\arccos\left(\displaystyle\frac{b+c+
\frac{2\xi}{\widetilde{E}+\widetilde{m}}}{b-c}\right)}{\sqrt{ad\left(b+\frac{\xi}{\widetilde{E}+\widetilde{m}}\right)
\left(c+\frac{\xi}{\widetilde{E}+\widetilde{m}}\right)}} \nonumber\\
\displaystyle &+&O\left(\frac{\sigma\gamma}{\widetilde{E}^2}\right).\label{eq14}
\end{eqnarray}
Adding expansions (\ref{eq10}) and (\ref{eq14}) and combining terms of like orders in $\sigma$, we obtain the transcendental equation determining the energy spectrum from (\ref{3a}),
\begin{eqnarray}
&&\displaystyle\frac{\eta_1\sqrt{\widetilde{E}^2-\widetilde{m}^2}}
{2\sigma(2\lambda-1)}-\eta\left(\frac{\eta_2^2}
{2\sigma(2\lambda-1)}+\lambda\, \xi\,\right)\nonumber\\
&&-\displaystyle\frac{\widetilde{E}\xi}{\sqrt{\widetilde{E}^2-\widetilde{m}^2}}
\log\left(\frac{\sigma\,\eta_2 \theta}
{4\,e\,(\widetilde{E}^2-\widetilde{m}^2)^2}\right)-\gamma\arccos\left(\displaystyle\frac{-\widetilde{E}\xi}{\theta}\right)\nonumber\\
&&\displaystyle
-\frac{\mbox{sgn}\,k}{2}\,\arccos\left(\displaystyle\frac{-\widetilde{m}\xi}
{\theta}\right)=\left(n_r+\displaystyle\frac{1}{2}\right)\pi.\label{eq15}
\end{eqnarray}
where
\begin{equation}\label{eta}
\eta=(1-2\lambda)^{-1/2}\arccos(\eta_1/\eta_2).
\end{equation}

Although Eq. (\ref{eq15}) is much simpler than ``exact'' quasiclassical equation (\ref{5}) for the energy levels,
solving it still requires numerical calculations. Below, we consider several limiting cases where Eq. (\ref{eq15}) is
simplified and can be investigated analytically.

For the parameter values $\sigma\lesssim 0.2\,\mbox{GeV}^2$ and $0.3<\xi<0.8$, the condition $\widetilde{E}\gg \widetilde{m}$ is well satisfied for all possible values of the level energies $E_{n_r k}$ of heavy-light mesons. If we expand the left-hand side of (\ref{eq15}) in $\widetilde{m}/\widetilde{E}\ll 1$ up to third-degree terms, we obtain the transcendental equation for $E_{n_r k}$:
\begin{eqnarray}
&&\displaystyle\left[(1-\lambda)A-\lambda\right]\widetilde{E}^2+2\widetilde{m}\widetilde{E}(1-\lambda)(\lambda A-1)\nonumber\\
&&\displaystyle-2\sigma(1-2\lambda)\left(\pi N +\xi\log\frac{\sigma|k|(1-\lambda)}{4\widetilde{E}^2}\right)\nonumber\\
&&\displaystyle-\lambda\widetilde{m}^2+\lambda[\lambda\widetilde{m}^2-2\sigma\xi(1-\lambda)]A=0,\label{eq15c}
\end{eqnarray}
where
\begin{equation}\label{eq15cc}
\begin{array}{c}
A=\displaystyle\frac{\arccos\left(\frac{\lambda}{1-\lambda}\right)}{\sqrt{1-2\lambda}},\\
\hspace{-3mm}N=\displaystyle n_r+\frac{1}{2}+\frac{\mbox{sgn}\,k}{4}+\frac{1}{\pi}
\left(\gamma\arccos\left(-\frac{\xi}{|k|}\right)-\xi\right).
\end{array}
\end{equation}
Solving this equation by the method of consecutive iterations, we obtain the desired expression for the
eigenvalues $E_{n_r k}$ in the first approximation (up to terms of the order $O(\sigma\gamma/\widetilde{E}^2)$):
\begin{eqnarray}
\displaystyle E^{\mbox{\footnotesize WKB(as)}}_{n_r k}&=&\zeta^{-1}\Biggl\{B+\Bigl(B^2+\zeta\Bigl[2\sigma(1-2\lambda)\Bigr.\Bigr.\Biggr.\nonumber\\
\displaystyle &\times&\Bigl(\xi\log\frac{\sigma|k|(1-\lambda)}{\left.4\widetilde{E}^{(0)}\right.^{2}}+3\xi+\lambda\xi A+\pi N\Bigr)\nonumber\\
\displaystyle&+& \Biggl.\Bigl.\Bigl.\lambda\widetilde{m}^2(1-\lambda A)\Bigr]\Bigr)^{1/2}\Biggr\}+\lambda V_0,\label{eq16}
\end{eqnarray}
where
\[\begin{array}{c}
\displaystyle\zeta=(1-\lambda)^2
A-\lambda-\frac{2\sigma\xi(1-2\lambda)}{\left.\widetilde{E}^{(0)}\right.^2},\\
\displaystyle B=(1-\lambda)(1-\lambda
A)\widetilde{m}-\frac{4\sigma\xi(1-2\lambda)}{\widetilde{E}^{(0)}},
\end{array}\]
and $\widetilde{E}^{(0)}=E^{(0)}-\lambda V_0$. Here, $E^{(0)}$ is the zeroth approximation for the energy on which the quantity $E_{n_r k}$ depends rather weakly, and we can set $E^{(0)}\approx E_{n_r k}(\xi)|_{\xi=0}$ in most cases.

We have obtained formula (\ref{eq16}) for the energy levels $E_{n_r k}$, which depend nonanalytically on the string
tension $\sigma$ and which therefore cannot be obtained in the perturbation theory framework. We mention that
for a purely scalar confinement ($\lambda=0$), formula (\ref{eq16}) is simplified to
\begin{eqnarray} \label{eq17}
&&E^{\mbox{\footnotesize WKB(as)}}_{n_r
k}\nonumber\\
&&=\frac{2}{\pi}\left[m+\sqrt{m^2+
\sigma\pi\left(\xi\log\frac{\sigma|k|}{(2 E^{(0)})^2}+\pi
N\right)}\right].
\end{eqnarray}

The results of calculating the energy levels $E^{\mbox{\footnotesize WKB}}_{n_r k}$ and $E^{\mbox{\footnotesize
WKB(as)}}_{n_r k}$ based on transcendental equation (\ref{5}) and asymptotic formula (\ref{eq16}) together with the exact values of $E_{n_r k}$ obtained by solving the Dirac equation numerically are presented in Table~\ref{tab1} for $n_r=0, 1, 2$ and $k=\pm 1,\pm 2$. In these calculations, we set the values of $\alpha_s$, $\lambda$,
$V_0$, $m_{u,d}$, and $m_{s}$ to those used in QCD to describe the states of $B$($b\overline{u}$ or $b\overline{d}$) and $B_{s}(b\overline{s})$ mesons. As can be seen in Table~\ref{tab1}, the quasiclassical values $E^{\mbox{\footnotesize WKB}}_{n_r k}$ and $E^{\mbox{\footnotesize WKB(as)}}_{n_r k}$ ensure the respective 1\% and 2\%
accuracies (except the energy of states with the radial quantum number $n_r=0$, for which the accuracy of
both formulas is about 8\%). The accuracy of determining $E_{n_r k}$ from quasiclassical formula (\ref{eq16}) is therefore such that the first-order approximation usually suffices for practical purposes.
\begin{table*}
\caption{{\footnotesize The results of calculating the level energies $E^{\mbox{WKB}}_{n_r k}$ (based on transcendental equation (\ref{5})) and $E^{\mbox{\footnotesize WKB(as)}}_{n_r k}$ (based on quasiclassical expression (\ref{eq16})) and also the exact values of $E_{n_r k}$ calculated at the parameter values $\alpha_s=0.3$, $\lambda=0.3$,
$V_0=-0.45$\,GeV and $m_{u,d}=0.33$\,GeV, $m_{s}=0.5$\,GeV (the energies are measured in GeV).}}\label{tab1}
\begin{center}
\begin{tabular}{|c|c|c|c|c||c|c|c|} \hline
\multicolumn{5}{|c||}{$b\overline{u}$,\quad
$b\overline{d}$}&\multicolumn{3}{|c|}{$b\overline{s}$}\\
\hline \multicolumn{2}{|c|}{$L_j$\,($n_r, k$)}&$E_{n_r
k}$&$E^{\mbox{{\footnotesize WKB}}}_{n_r
k}$&$E^{\mbox{{\footnotesize WKB(as)}}}_{n_r k}$&
$E_{n_r k}$&$E^{\mbox{{\footnotesize WKB}}}_{n_r k}$&$E^{\mbox{{\footnotesize WKB(as)}}}_{n_r k}$\\
\hline &(0, -1)&0.4327&0.4408&0.4729&0.5248&0.5322&0.5623\\
\cline{2-8}$S_{1/2}$&(1, -1)&0.8796&0.8838&0.8943&0.9750&0.9791&0.9912\\
\cline{2-8}&(2, -1)&1.1978&1.2009&1.2066&1.2946&1.2976&1.3049\\
\hline&(0, -2)&0.7355&0.7373&0.7504&0.8376&0.8392&0.8460\\
\cline{2-8}$P_{3/2}$&(1, -2)&1.0880&1.0892&1.0947&1.579&1.590&1.1927\\
\cline{2-8}&(2, -2)&1.3658&1.3667&1.3699&1.4650&1.4659&1.4685\\
\hline &(0, 1)&0.7249&0.7293&0.7030&0.8235&0.8278&0.7985\\
\cline{2-8}$P_{1/2}$&(1, 1)&1.0701&1.0733&1.0594&1.1696&1.1728&1.1572\\
\cline{2-8}&(2, 1)&1.3470&1.3496&1.3405&1.4466&1.4492&1.4390\\
\hline &(0, 2)&0.9661&0.9671&0.9343&1.0655&1.0665&1.0315\\
\cline{2-8}$D_{3/2}$&(1, 2)&1.2588&1.2596&1.2385&1.3583&1.3591&1.3369\\
\cline{2-8}&(2, 2)&1.5058&1.5066&1.4914&1.6052&1.6059&1.5901\\
\hline
\end{tabular}
\end{center}
\end{table*}

\begin{figure}[hbtp]
\includegraphics*[scale=0.65]{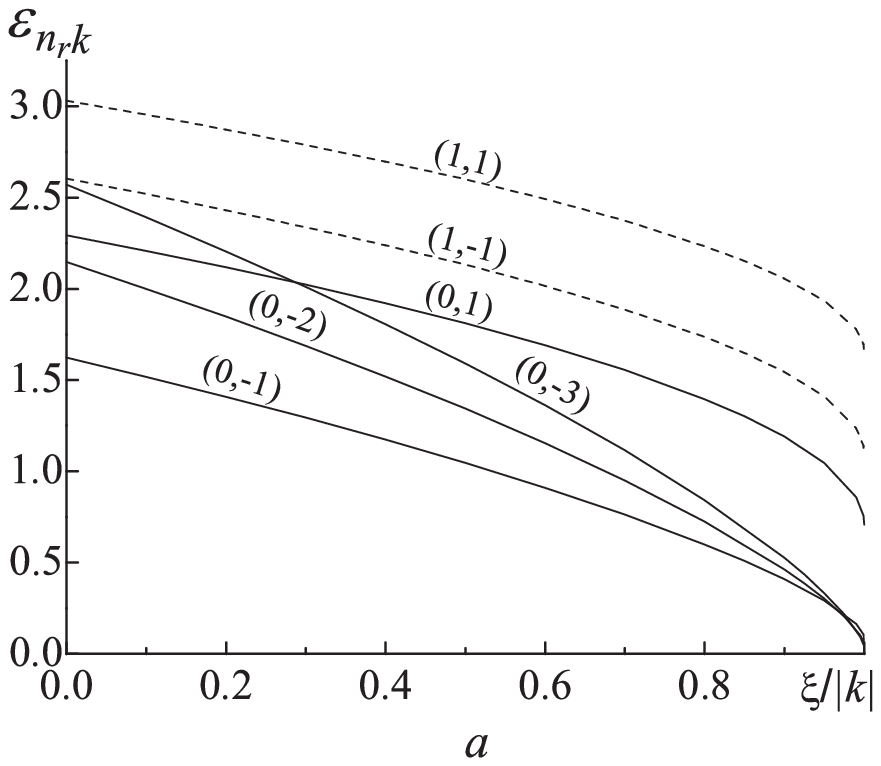}
\includegraphics*[scale=0.65]{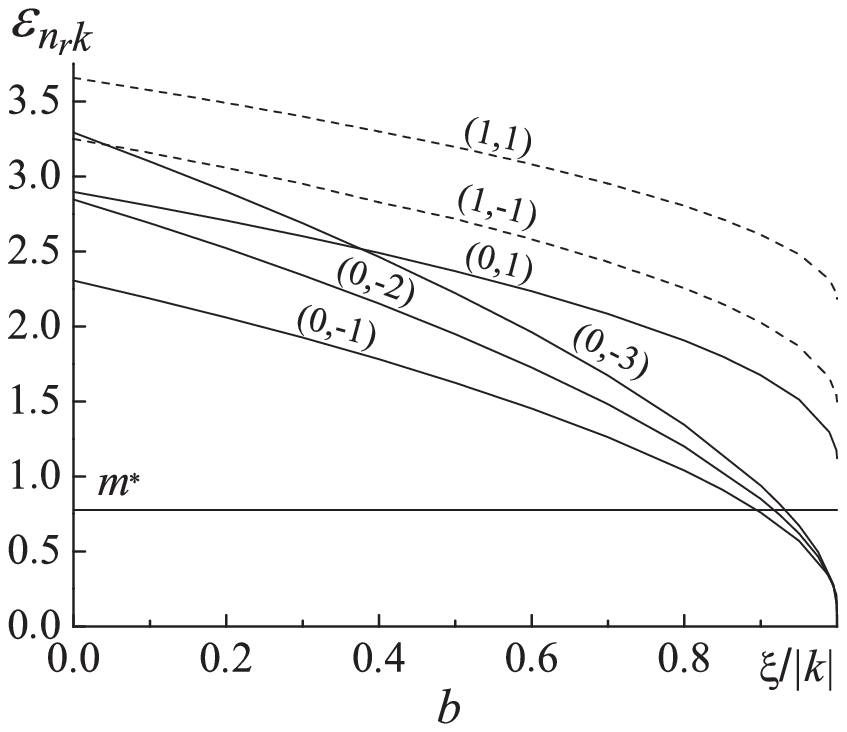}
\includegraphics*[scale=0.65]{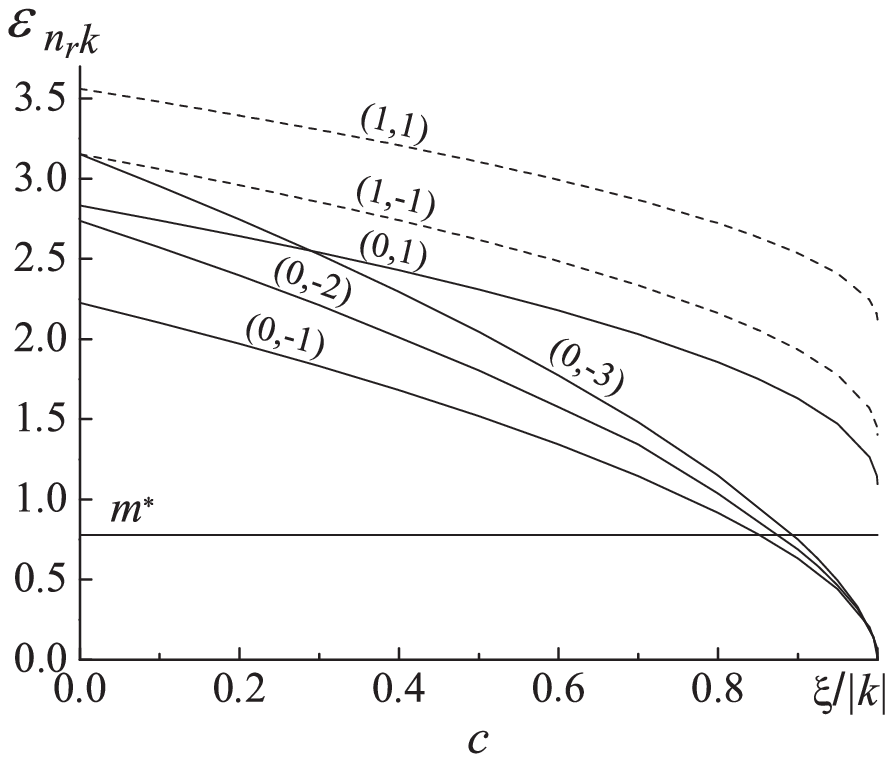}
\includegraphics*[scale=0.65]{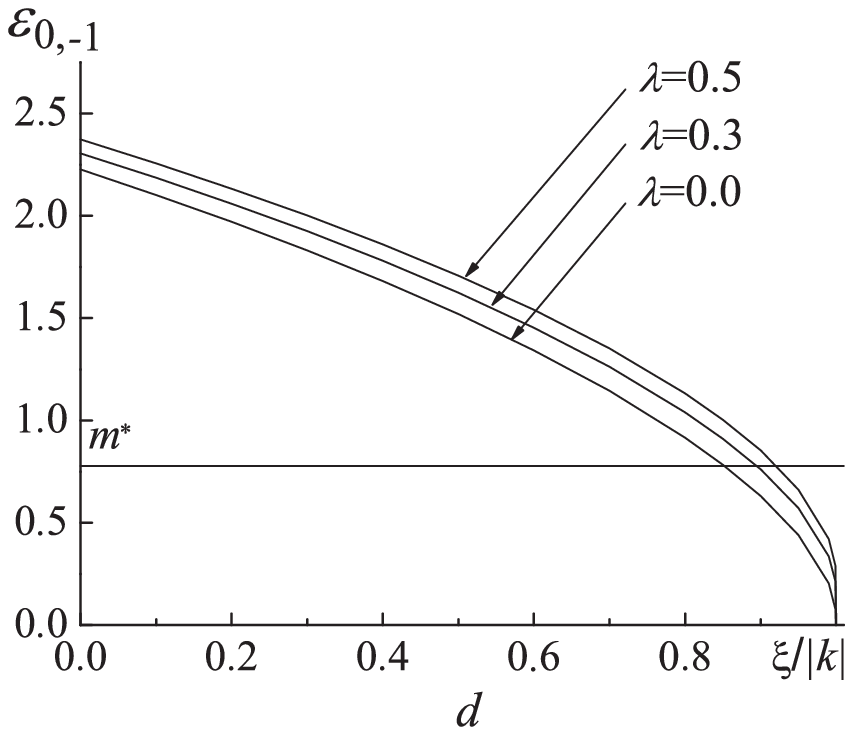}
\caption{(a--c) The dependence of the level energies $\varepsilon_{n_r k}=E_{n_r k}/\sqrt{\sigma}$ on $\xi/|k|$. Solid lines correspond to the lowest levels ($n_r=0$) with the given value of $k$, and dashed lines correspond to the excited states ($n_r=1$). The parameter values are (a) $m$=0 and $\lambda=0$, (b) $m=0.33\,\mbox{GeV}$ and $\lambda=0$,
(c) $m=0.33\,\mbox{GeV}$, $\sigma=0.18\, \mbox{Gev}^2$, and $\lambda=0.3$. Here, $m^*$=$m/\sqrt{\sigma}$. (d) The dependence of the level energy $\varepsilon_{0,-1}$ on $\xi/|k|$ for different values of the parameter $\lambda$ at $m$ = 0.33 GeV.}\label{f2}
\end{figure}

In order to find the dependence of the energy eigenvalues $E_{n_r k}$ on the Coulomb coupling constant $\xi$, we solved
transcendental equation (\ref{5}) numerically with the following choice of parameters determining the form
of initial interaction potentials (\ref{potential}): $\alpha_s=0.3$, $\lambda=0.3$, $V_0=-0.45$\,GeV, and $m_{u,d}=0.33$\,GeV. The graphs of dependences of energy levels on the ratio $\xi/|k|$ are shown in Fig.~\ref{f2}, where solid lines indicate the dependence of several lowest levels ($n_r=0$) with the given value of $k$ and dashed lines correspond to the excited states ($n_r=1$). As could be expected, as the Coulomb parameter $\xi$ increases, level energy decreases monotonically and develops a square-root singularity as $\xi\rightarrow |k|$. This is a manifestation of the ``falling to center'' phenomenon for the Dirac equation in composite field (\ref{potential}) with the vector potential $V(r)$, which has the Coulomb singularity at zero, $V(r)\approx V_{\mathrm{Coul}}(r)=-\xi/r$ as $r\rightarrow 0$. As is known (see Sec.~\ref{s3}), every cut-off of the potential $V(r)$ at small distances removes the square-root singularity in the energies $E_{n_r k}$, and the curve of the level of $E_{n_r k}(\xi)$ can then be smoothly continued into the domain $E<0$.

It can be seen from Fig.~\ref{f2} that for the states with the same $n_r$, the energy levels with $k>0$ lie much
higher than levels with $k<0$. This is the influence of the centrifugal barrier (for instance, this barrier is
absent for states with $k=-1$, while it suppresses the probability of the presence of the quark at large distances for states with $k=+1$). These conclusions are completely confirmed by numerically solving Dirac system (\ref{1}) with potentials (\ref{potential}); the results of this were presented in \cite{Mur4}.

We also note that the energies of the lowest levels ($n_r=0$) with $k<0$ reach the zero level ($E=0$) at the maximum possible value of the Coulomb coupling constant $\xi=-k$ (see Figs.~\ref{f2}a--\ref{f2}c). All other states
also have the singularity of the square-root type at $\xi=|k|$, but their energies remain positive.

The above study of the spectrum of Dirac equation in composite field (\ref{potential}) using the WKB approximation
is of practical interest because calculating integrals in quantization condition (\ref{3a}) is much easier in many cases than finding exact values of energy levels by numerically solving system of radial Dirac equations (\ref{1}).

\section{The mass spectrum of heavy-light quark systems}\label{s5}

The qualitative picture of forming bound states in a $Q\bar{q}$ system is determined by the presence of the
scale parameter $\Lambda_{\mbox{\footnotesize QCD}}$ of the confinement of the light antiquark $\bar{q}$: $\Lambda_{\mbox{\footnotesize QCD}}\ll m_{Q}$, where $m_Q$ is the mass of the heavy quark $Q$. Under this condition, the heavy quark $Q$ affects the light quark $\bar{q}$ as a local static source of the color (gluon) QCD field. The presence of a small parameter $\Lambda_{\mbox{\footnotesize QCD}}/m_{Q}\ll 1$ allowed developing
powerful means for studying QCD in interactions between heavy and light quarks. For example, a consistent
scheme of the effective theory of heavy quarks for hadronic systems with one heavy quark ($Q\bar{q}$, $Qqq$) was
developed (see, e.g., \cite{Matsuki} and the references therein). In the leading term of this theory (i.e., in the static limit as $m_{Q}\rightarrow\infty$), first, the spin of the heavy quark $Q$ splits from the interaction with weakly virtual gluons, second, the effective Hamiltonian exactly corresponds to the Dirac Hamiltonian of one-particle problem (\ref{1}), and the energy of the spin.orbital interaction of the light antiquark $\bar{q}$ becomes the leading term of spin interactions. This is manifested in the approximate Isgur-Wise spin symmetry \cite{Isgur} for the heavy quark.

In the leading order in $1/m_{Q}$, the mass spectrum of meson states with one heavy quark is given by the
expression \cite{Matsuki,Ebert,Khr,Lichtenberg}
\begin{equation}\label{eq5}
M^{\mbox{\footnotesize theor}}_{n_r k}(Q\bar{q})=E_{n_r
k}+\sqrt{E^{2}_{n_r k}-m^{2}_q+m^{2}_Q},
\end{equation}
where $m_Q$ and $m_q$ are the masses of the heavy quark $Q$ and the light quark $\bar{q}$ constituting the $Q\bar{q}$ meson. Calculating the mass spectrum of $Q\bar{q}$ mesons therefore reduces to consistently calculating the energy
eigenvalues of Dirac equation (\ref{1}) in composite field (\ref{potential}) whose source here is the heavy quark $Q$.

The symmetry properties of Dirac equation (\ref{1}) drastically simplify the problem of classifying states of
heavy-light mesons. Because the Hamiltonian of Eq. (\ref{1}) does not contain terms describing the interaction
of the spin of the $Q$ quark with the orbital and spin moments $\vec{l}$ and $\vec{s}_q$ of the light antiquark, both the spin moment $\vec{S}_Q$ of the heavy quark $Q$ and the total moment $\vec{j}=\vec{s}_q+\vec{l}$ of the light antiquark $\bar{q}$ are two separate integrals of motion. This allows classifying the states by the quantum numbers $j=\frac{1}{2}, \frac{3}{2},\ldots$ of the operator of the total moment of the light antiquark $\bar{q}$, while the states of the total moment of the composite $Q\bar{q}$ system $\vec{J}=\vec{j}+\vec{S}_Q$ are degenerate with respect to the orientation of the spin $\vec{S}_Q$ of the heavy quark $Q$. Two almost degenerate states of the composite $Q\bar{q}$-system with $J=j\pm 1/2$ in the spin symmetry approximation \cite{Isgur} therefore correspond to each state of the Dirac equation with the given j and with the spatial parity $P=(-1)^{l+1}$. Masses of the $j^P$-states of the $Q\bar{q}$ meson are also degenerate with respect to $J$, and these states therefore have identical wave functions.

The values $l=0$ ($\mathrm{S}$ states in the quark-antiquark model) and $j=1/2^-$ correspond to the ground state of the $Q\bar{q}$ meson. This doublet consists of two states $J^P=(0^-, 1^-)$. In the case $l=1$ (the $\mathrm{P}$ state
in the quark model), we have two states with $j=1/2^+$ and $j=3/2^+$ and two corresponding doublets $J^P=(0^+,
1^+)$ and $J^P=(1^+, 2^+)$.

As usual, we introduce a concise notation for the families of $D$ and $D_s$ mesons: ($D^*_0$, $D'_1$) are the components of the charmed doublet $J^P=(0^+, 1^+)$ with $j=1/2^+$ for nonstrange states (the $c\bar{u}$ system),
($D^*_{s0}$, $D'_{s1}$) are the components of the same doublet for strange states (the $c\bar{s}$ system), and ($D_1$,
$D^*_2$) and ($D_{s1}$, $D^*_{s2}$) are the components of the doublet $J^P=(1^+,2^+)$ with $j=3/2^+$ for the respective nonstrange and strange states. We also use the analogous notation system for $B$ and $B_s$ families.

Above, we did not take the level hyperfine structure (HFS) into account, and the proposed potential
model can predict only the position of the center of masses of the HFS multiplet comprising sublevels with
different moments $\vec{J}=\vec{j}+\vec{S}_Q$. In actual $Q\bar{q}$ systems, the degeneracy of doublet states corresponding to different moments $J=j\pm 1/2$ at the given $j$ is broken primarily because of the $\vec{s}_q\,\vec{S}_Q$ interaction. Therefore, to be able to compare our theoretical predictions with experimental data, we present the observation values for the centers of masses of the HFS multiplets in Tables~\ref{D-meson}--\ref{Bs-meson}; these centers of masses are calculated by the
known formula
\begin{equation}\label{eq19}
M_{\mbox{exp}}=\frac{\sum \limits_J\left((2J+1)\,
M_J\right)}{\sum \limits_J(2J+1)},
\end{equation}
where $M_J$ is the experimental value of the mass of state with the given $J$.

Based on these observations, we have tried to describe the spectra of masses of low-lying states of the
heavy-light $B(b\bar{u}\,\mbox{or}\, b\bar{d})$, $B_{s}(b\bar{s})$, $D(c\bar{u}\, \mbox{or}\, c\bar{d})$, and $D_{s}(c\bar{s})$ mesons considering $\sigma$ and $\lambda$ to be universal quantities and setting the values of the parameters $\alpha_s$ and $V_0$ constant in every family of heavy-light mesons allowing them to vary slightly only when passing from one family to another. All the parameters $\sigma$, $\lambda$, $\alpha_s$, and $V_0$ of potential model (\ref{potential}) were determined by fitting the known data for the mass spectra of
pseudoscalar $D$ and $B$ mesons. The found values of the parameters are consequently used below in other
applications in the framework of our approach, for example, when describing the spectra of the strange $D_s$ and $B_s$ mesons.

We use only one a priori restriction: the value of the coefficient $\lambda$ of mixing between the vector and
scalar long-range potentials $V_{\mathrm{conf}}(r)$ and $S_{\mathrm{conf}}(r)$ must lie in the interval $0\leqslant\lambda<1/2$ for the EP $U(r,E)$ of interaction model (\ref{potential}) to be a confining-type potential. The value of the parameter $\lambda$ was obtained by fitting experimental data \cite{Experimemt,Godfrey2} on the fine structure of $\mathrm{P}$-wave levels in $D$ and $B$ mesons. It was established that the fine structure of the $\mathrm{P}$-wave states in the heavy-light ($D$, $D_s$, $B$, and $B_s$) mesons is primarily sensitive
to the choice of the mixing coefficient $\lambda$ and to the value of the strong coupling constant $\alpha_s$. Comparing the results of calculations based on formulas (\ref{5}) and (\ref{eq5}) with the experimental data \cite{Experimemt,Godfrey2}, we find that the best agreement is reached at $\lambda=0.3$ and for the parameter choices
\begin{eqnarray*}
&\sigma=0.18\,\mbox{GeV}^2,\, \alpha_s(c\bar{u}\,
\mbox{or}\, c\bar{d})=0.386,\, \alpha_s(b\bar{u}\,
\mbox{or}\, b\bar{d})=0.3, &\\
&V_0(c\bar{u}\, \mbox{or}\, c\bar{d})=-375\,\mbox{MeV},\phantom{x} V_0(b\bar{u}\, \mbox{or}\,
b\bar{d})=-450\,\mbox{MeV}.&
\end{eqnarray*}
For the masses of $u$, $d$, $s$, $c$, and $b$ quarks, we used their constituent masses $m_{u,d}=330$\,MeV, $m_{s}=500$\,MeV, $m_{c}=1550$\,MeV, and $m_{b}=4880$\,MeV. When calculating the mass spectrum, we neglected electromagnetic interaction and the difference of the masses of $u$ and $d$ quarks, therefore considering the particles $D^+$, $D^-$, $D^0$, and $\bar{D}^0$, for example, to be the same state of the $Q\bar{q}$ system, $J^P=0^-$. Correspondingly, we do not distinguish between the interaction parameters $\sigma$, $\lambda$, $\alpha_s$, and $V_0$ for these particles. The mass spectra of $D$ and $D_s$ mesons calculated in this approximation and by means of numerical solutions of system (\ref{1}) are presented in Tables ~\ref{D-meson} and \ref{Ds-meson}.
\begin{table}[h]
\caption{{\footnotesize The mass spectrum and the mean radii of $D$ mesons
obtained in the WKB approximation and numerically for potentials (\ref{potential})
(masses are expressed in MeV and the mean radii are expressed
in Fm).}}\label{D-meson}
\begin{center}
\begin{tabular}{|c|c|c|c|c|c|c|}
\hline\multicolumn{2}{|c|}{$L_j$\,($n_r,
k$)}&$M_{\mbox{{\footnotesize num}}}$&$M_{\mbox{{\footnotesize WKB}}}$&$M_{\mbox{{\footnotesize
exp}}}$&$\langle r\rangle_{\mbox{{\footnotesize num}}}$ &$\langle
r\rangle${\footnotesize(\ref{7})}\\
\hline $S_{1/2}$&(0, -1)&1989.1&2001.5&1971.1&0.472&0.402\\
\cline{2-7}&(1, -1)&2624.5&2632.3&$<$\,2637&0.684&0.664\\
\hline $P_{3/2}$&(0, -2)&2440.1&2443.2&2447.3&0.678&0.632\\
 \cline{2-7}&(1, -2)&2979.7&2981.9&--&0.856&0.833\\
\hline $P_{1/2}$&(0, 1)&2395.2&2403.7&2407.8&0.513&0.568\\
\cline{2-7} &(1, 1)&2926.8&2933.4&--&0.770&0.788\\
\hline
\end{tabular}
\end{center}
\end{table}
\begin{table*}
\caption{{\footnotesize The mass spectrum and the mean radii of $D_s$ mesons
obtained in the WKB approximation and numerically for potentials (\ref{potential})
(masses are expressed in MeV and the mean radii are expressed
in Fm).}}\label{Ds-meson}
\begin{center}
\begin{tabular}{|c|c|c|c|c|c|c|c|}
\hline\multicolumn{2}{|c|}{$L_j$\,($n_r,
k$)}&$M_{\mbox{{\footnotesize num}}}$&$M_{\mbox{{\footnotesize WKB}}}$&\multicolumn{2}{c|}{$M_{\mbox{{\footnotesize exp}}}$}&
$\langle r\rangle_{\mbox{{\footnotesize num}}}$&$\langle r\rangle${\footnotesize (\ref{7})}\\
\hline $S_{1/2}$&(0, -1)&2057.2&2069.0&\multicolumn{2}{c|}{2072}&0.416&0.359\\
\cline{2-8}&(1, -1)&2729.4&2737.4&\multicolumn{2}{c|}{--}&0.646&0.628\\
\hline$P_{3/2}$&(0, -2)&2550.1&2552.1&2559.2\,(I)&2530.7\,(II)&0.625&0.588\\
 \cline{2-8}&(1, -2)&3105.2&3107.2&--&--&0.814&0.795\\
\hline $P_{1/2}$&(0, 1)&2499.7&2508.5&2423.8\,(I)&2480.9\,(II)&0.504&0.536\\
\cline{2-8}&(1, 1)&3051.7&3058.5&--&--&0.739&0.756\\
\hline
\end{tabular}
\end{center}
\end{table*}

The agreement between the numerical result (result of WKB approximation) and experiment is less than 1\% (1.5\%), except for the masses of states
$\mathrm{P}_{3/2}$ and $\mathrm{P}_{1/2}$ of the $c\bar{s}$ system for which the mismatch depends on the interpretation of the $D_{s1}$(2536)$^{\pm}$ meson and is 3.1\% (3.5\%) if we consider it to be the vector state $J^P=1^+$ belonging to the doublet $j=3/2^+$ or 0.4\% (0.3\%) if we consider it to be the state $J^P=1^+$ of the doublet $j=1/2^+$. There is a rather broad
spectrum of opinions concerning the identification of the states $\mathrm{P}_{3/2}$ and $\mathrm{P}_{1/2}$ of the meson with the quark content $c\bar{s}$ (see, e.g., \cite{Abe,Link,Anderson,Besson,Aubert,Colangelo,Bardeen,Godfrey1,Colangelo1,Cahn,Lucha}). For example, the state $J^P=2^+$ of a relatively narrow doublet $j=3/2^+$ is related to $D_{s2}$(2573), while the vector state $J^P=1^+$ belonging to the doublet $j=3/2^+$ is generally
related to the isotopic singlet $D_{s1}$(2536)$^{\pm}$ meson with the mass 2535.35$\pm$0.34$\pm$0.5 MeV (values (I) for $M_{\mbox{exp}}$ in Table~\ref{Ds-meson}) \cite{Abe,Link,Anderson,Besson,Aubert,Colangelo,Bardeen,Godfrey1,Colangelo1}. On the other hand, the state $D_{s1}$(2536)$^{\pm}$ was associated with the state $J^P=1^+$ of the wide doublet $j=1/2^+$ in \cite{Cahn,Lucha} (values (II) in Table~\ref{Ds-meson}). Therefore, a reliable experimental identification of this state is still lacking. We note that our calculations agree better with the second possibility.

For $b\bar{u}$ and $b\bar{s}$ systems, we obtained a good agreement of our results with the experimental data for
the ground state with $j=1/2^-$ and for the $\mathrm{P}$ state with $j=3/2^+$ (see Tables~\ref{B-meson} and \ref{Bs-meson}). For states in the doublet $j=1/2^+$, we have only theoretical predictions of other authors. For the $b\bar{u}$ system, our results agree with the data obtained in \cite{Sadzikowski}, and a remarkable agreement with the results in \cite{Colangelo,Bardeen,Deandrea} was obtained for the $b\bar{s}$ system.

\begin{table}[h]
\caption{{\footnotesize The mass spectrum and the mean radii of $B$ mesons
obtained in the WKB approximation and  numerically for potentials (\ref{potential})
(masses are expressed in MeV and the mean radii are expressed
in Fm).}}\label{B-meson}
\begin{center}
\begin{tabular}{|c|c|c|c|c|c|c|}
\hline \multicolumn{2}{|c|}{$L_j$\,($n_r,
k$)}&$M_{\mbox{{\footnotesize num}}}$&$M_{\mbox{{\footnotesize WKB}}}$&$M_{\mbox{{\footnotesize
exp}}}$&$\langle r\rangle_{\mbox{{\footnotesize num}}}$
&$\langle
r\rangle${\footnotesize(\ref{7})}\\
\hline
$S_{1/2}$&(0, -1)&5320.7&5329.5&5313.5&0.516&0.448\\
\cline{2-7}&(1, -1)&5827.3&5832.2&--&0.728&0.708\\
\hline$P_{3/2}$&(0, -2)&5659.6&5661.6&$<$\,5698&0.711&0.666\\
\cline{2-7}&(1, -2)&6076.9&6078.4&--&0.888&0.865\\
\hline&(0, 1)&5647.4&5652.4&5751.6\,\cite{Colangelo}&0.577&0.612\\
$P_{1/2}$&&&&5624\,\cite{Sadzikowski}&&\\
\cline{2-7}&(1, 1)&6055.1&6059.0&--&0.812&0.829\\
\hline
\end{tabular}
\end{center}
\end{table}

\begin{table}[h]
\caption{{\footnotesize The mass spectrum and the mean radii of $B_s$ mesons
obtained in the WKB approximation and  numerically for potentials (\ref{potential})
(masses are expressed in MeV and the mean radii are expressed
in Fm).}}\label{Bs-meson}

\begin{center}
\begin{tabular}{|c|c|c|c|c|c|c|}
\hline \multicolumn{2}{|c|}{$L_j$\,($n_r,
k$)}&$M_{\mbox{{\footnotesize num}}}$&$M_{\mbox{\footnotesize WKB}}$&$M_{\mbox{{\footnotesize
exp}}}$&$\langle r\rangle_{\mbox{{\footnotesize num}}}$
&$\langle
r\rangle${\footnotesize(\ref{7})}\\
\hline
$S_{1/2}$&(0, -1)&5407.4&5415.6&5404.8&0.457&0.404\\
\cline{2-7}&(1, -1)&5926.3&5931.2&--&0.688&0.671\\
\hline$P_{3/2}$&(0, -2)&5763.7&5765.6&$<$\,5853&0.656&0.619\\
\cline{2-7}&(1, -2)&6185.5&6186.8&--&0.845&0.826\\
\hline&&&&5751.8\,\cite{Colangelo}&&\\
&&&&5753.3\,\cite{Bardeen}&&\\
$P_{1/2}$&(0, 1)&5747.2&5752.2&5700.5\,\cite{Sadzikowski}&0.547&0.575\\
&&&&5755.0\,\cite{Deandrea}&&\\
&&&&5790.3\,\cite{Green}&&\\
\cline{2-7}&(1, 1)&6162.8&6166.8&--&0.779&0.795\\
\hline
\end{tabular}
\end{center}
\end{table}

In the leading approximation (in $1/m_{Q}$), the wave functions and excitation energies of the strange quark in the field of a heavy $c$ or $b$ quark reproduce the corresponding characteristics of heavy-light mesons with light $u$ and $d$ quarks with high accuracy. Therefore, up to an additive upward shift of masses on the value of the current mass of the strange quark
\[m_s\approx M[D_s]-M[D]\approx M[B_s]-M[B]\approx0.1\,\mbox{GeV}\]
the level systems for $D_s$ and $B_s$ mesons coincides with the respective level systems for $D$ and $B$ mesons if we do not take the level splitting depending on the spin of the heavy quark into account. Further, the spin-orbital splitting of lower states of $D_s$ and $B_s$ mesons for the levels $\mathrm{P}_{3/2}$ and $\mathrm{P}_{1/2}$ is 35\% larger than that of the $D$ and $B$ mesons.

Not only the spectrum of bound systems, all other observable characteristics of heavy-light mesons can be calculated in the framework of the quasiclassical approach under consideration. For example, an important meson characteristics is the mean radius $\langle r\rangle$, which determines the radius of the light quark orbit in a definite state $|n_r k \rangle$ in the case of hydrogen-like quark systems. We first obtain general formulas expressing the means of type
$\left\langle r^m\right\rangle$ (i.e., the moments of the probability distribution density) in terms of quasiclassical asymptotic expressions for solutions of the Dirac equation. Using the standard procedure, we obtain the known quasiclassical formula
\begin{eqnarray}
\displaystyle\left\langle r^m\right\rangle&=&\int\limits_{0}^{\infty}\chi^+ r^m \chi\,
dr=\int\limits_{0}^{\infty}(\left|F\left(r\right)\right|^2
+\left|G\left(r\right)\right|^2)r^m dr \nonumber\\
\displaystyle&\approx&\frac{2}{T}\int\limits_{r_0}^{r_1}\frac{E-V\left(r\right)}{p\left(r\right)}
\, r^m\,dr,\label{6}
\end{eqnarray}
where the period $T$ of radial oscillations of the classical relativistic particle is given by the formula $T=2\int^{r_{1}}_{r_{0}}(E-V(r))/p(r)\,dr$ \cite{Lazur}.

All the integrals in (\ref{6}) can be expressed in terms of complete elliptic integrals (A.1). In particular,
the mean radius of the bound state is
\begin{equation}\left\langle
r\right\rangle =\frac{4\left[n_1 F\left(\chi\right)+n_2
E\left(\chi\right) + n_3\Pi\left(\nu,\chi\right)\right]}
{T\sqrt{(a-c)(b-d)(1-2\lambda)}\,|\sigma|},\label{7}
\end{equation}
\begin{equation}
T =\frac{4\left[n_4 F\left(\chi\right)+n_5 E\left(\chi\right) +
n_6\Pi\left(\nu,\chi\right)\right]}
{\sqrt{(a-c)(b-d)(1-2\lambda)}\,|\sigma|}\label{7a}
\end{equation}
where the quantities $n_i$ ($i=1,\ldots, 6$) are defined in Appendix. The calculation results for $\langle r\rangle$, according to formulas (\ref{7}) and (\ref{7a}) for different states of $D$, $D_s$, $B$, and $B_s$ mesons are presented in the last columns in Tables~\ref{D-meson}--\ref{Bs-meson}. We see that the quasiclassical approximation well describes the numerical simulation results $\langle r\rangle_{\mbox{{\footnotesize num}}}$ and ensures an accuracy up to 3\% (except the ground state). Calculations demonstrate that the mean radius of the $Q\bar{q}$ system increases monotonically as the energy increases.

In addition to the ``exact'' quasiclassical formulas (\ref{7}) and (\ref{7a}), it is desirable to find approximate analytic expressions for the quantities $\langle r\rangle$ and $T$. We already addressed an analogous problem in the
preceding section when constructing asymptotic approximations for the quantization integrals.

If the condition $\sigma/\xi\widetilde{m}^2\ll1$ is satisfied in the spectral domain $\widetilde{E}<\widetilde{m}$, then the only essential contribution to the integral determining the mean radius $\langle r\rangle$ comes from the domain of the integration variable $r$ where the long-range potential $v(r)$ can be considered a small perturbation. Neglecting this potential, we obtain the expressions for the mean radius and the period in the zeroth approximation:
\begin{equation}\label{8b}
\langle r\rangle\approx\frac{\pi\widetilde{E}_0}{T\mu_0^3}
\left(\frac{3\xi^2\widetilde{m}^2}{\mu_0^2}-k^2\right),\quad
T\approx\frac{2\pi\xi\widetilde{m}^2}{\mu_0^3}.
\end{equation}
A more accurate expression (than (\ref{8b})) for the mean radius can be obtained if we use the exact solutions of
Dirac system (\ref{1}) in the Coulomb field in the integral $\int_{0}^{\infty}(\left|F\left(r\right)\right|^2
+\left|G\left(r\right)\right|^2)r\,dr \equiv \langle r\rangle$ \cite{Ahiezer}. The resulting expression for the mean radius of the hydrogen-like system becomes
\begin{equation}\label{8c}
\langle r\rangle_{\mathrm{Coul}}=\frac{\widetilde{E}_0}
{2\xi\widetilde{m}^2}\left(\frac{3\xi^2\widetilde{m}^2}{\mu_0^2}
-k^2-\frac{k\widetilde{m}}{\widetilde{E}_0}\right)
\end{equation}
and it coincides with (\ref{8b}) at large values of the radial quantum number $n_r$.

This simple approximation ensures an amazingly good accuracy for deeply lying levels (but, of course, not at $E=0$). For example, for the first three terms $1\mathrm{S}_{1/2}$, $1\mathrm{P}_{1/2}$, and $2\mathrm{S}_{1/2}$ of the $b$ quark ($m_b=4.88$ GeV), we obtain the respective values $\langle r\rangle=$ 0.153 Fm, 0.501 Fm, and 0.609 Fm from (\ref{8c}), while the exact calculation (the numerical solution of the Dirac equation with potentials (\ref{potential}) at $\xi=0.4$, $\lambda=0.3$, $V_0=-0.45$ GeV, and $\sigma=0.18$ GeV$^2$) yields the respective values
$\langle r\rangle=$ 0.153~Fm, 0.493~Fm, and 0.600~Fm. Our approximation therefore ensures a high accuracy in the case of heavy quarks.

Unfortunately, the domain of applicability of such an approximation is restricted by the condition $\sigma/\xi\widetilde{m}^2\ll1$. Because the problem of a size of a bound state of the $Q\bar{q}$ system is important, we consider it from the quantitative standpoint. We use the fact that the condition $\sigma\gamma/\widetilde{E}^2\ll1$ is satisfied for all typical values of the parameters $\xi$ and $\sigma$ of heavy-light quarks in the spectrum domain
$\widetilde{E}>\widetilde{m}$ under investigation. In this case, the light quark motion is mainly determined by the linear potential, and the Coulomb interaction can be considered a perturbation. In some cases, the zeroth approximation suffices for calculating $\langle r\rangle$ and $T$,
\begin{eqnarray}\label{8}
\hspace{-4mm}\langle
r\rangle\approx\frac{2}{T\sigma^2(1-2\lambda)}\left\{\left(
\frac{3\lambda\,\eta_1}{2(1-2\lambda)}+\widetilde{E}\right)
\sqrt{\widetilde{E}^2-\widetilde{m}^2}\right. \nonumber \\
\left.-\frac{\arccos\frac{\eta_1}{\eta_2}}{\sqrt{1-2\lambda}}\left[\widetilde{E}
\eta_1+\frac{\lambda}{2}\left(\frac{3\eta_1^2}{1-2\lambda}
+\widetilde{E}^2-\widetilde{m}^2\right)\right]
\right\},
\end{eqnarray}
\begin{eqnarray}\label{8a}
T\approx\frac{2}{\sigma
(1-2\lambda)}\left[-\lambda\sqrt{\widetilde{E}^2-\widetilde{m}^2}
+(1-\lambda)\,\eta\,\eta_2\right],
\end{eqnarray}
where the quantities $\eta_1$ and $\eta_2$ are determined in (\ref{theta}), and the quantity $\eta$ is determined in (\ref{eta}). For example, for the first three terms $1\mathrm{S}_{1/2}$, $1\mathrm{P}_{1/2}$, and $2\mathrm{S}_{1/2}$ of the $B$ meson ($m_b=4.88$ GeV and $m_u=0.33$  GeV), we obtain the respective quantities $\langle r\rangle$ = 0.381~Fm, 0.576~Fm, and 0.681~Fm in approximation (\ref{8}), (\ref{8a}), and the calculation using ``exact'' formulas (\ref{7}) and (\ref{7a}) (at $\alpha_s=0.3$, $\lambda=0.3$, $V_0=-0.45$\,GeV, $\sigma=0.18$ Gev$^2$) yields the respective values 0.448~Fm, 0.612~Fm, and 0.708~Fm. This approximation therefore ensures an acceptable accuracy for calculating the mean radii
of the $Q\bar{q}$ mesons.

In contrast to the masses of bound states of the heavy-light mesons, the wave functions of the mixed mesons at the zero, which determine the leptonic constants and normalizations of cross-sections of formation $D$- and $B$-mesons, are more sensitive to global properties of initial potentials $S(r)$ and $V(r)$. We now turn to deriving these wave functions.

\section{Asymptotic coefficients of a wave function}\label{s6}

Asymptotic coefficients $C_{F,G}$ at the zero and $A_{F,G}$ at the infinity are the characteristic parameters of a wave function. We shall give simple analytical approximations for these coefficients which describe the results of numerical calculations quite well.

First of all we consider the construction rules of asymptotic expansions of solutions of the Dirac equation at zero ($r\rightarrow 0$) along with more standard expansions of solutions, when
$r\rightarrow\infty$. For the considered potentials (\ref{potential}) we have
\begin{equation}\label{9}
F,G=C_{F,G}\,r^{\gamma}+..., \, r\rightarrow0,\quad
C_{F}/C_{G}=(k-\gamma)/\xi,
\end{equation}
and for wave functions of the discrete spectrum (0$\leqslant\lambda <1/2$) the normalization condition
$\int\limits^{\infty}_{0}\left(F^2+G^2\right)dr=1$ is satisfied. Values of $F^2(0)$, $G^2(0)$ (or, more precisely, of $C^2_{F,G}$) define the probability to discover particles at small distances from one another and are of considerable physical interest especially in the case of systems, in which interactions of two various types (for example, the Coulomb interaction and long-range one $v(r)$) exist.

In the classically forbidden range $0\leqslant r<r_0$ the wave function of oscillating type is changed by the solution decreasing exponentially with increasing $r$ (see Fig.~\ref{f1}).
Matching the WKB-solutions of the Dirac equation on both sides of the turning point $r_0$, we obtain the quasiclassical expressions for the radial wave functions $F(r)$ and $G(r)$ in the classically forbidden region $0\leqslant r<r_0$:
\begin{eqnarray}\label{10}
F(r)=(-1)^{n_r}\frac{C_{1}^{\pm}}{2}&&\left[\displaystyle\frac{E-V+m+S}{q(r)}\right]^{1/2}\nonumber\\
&&\times\exp\left[-\int\limits^{r_{0}}_{r}\left(q-\frac{k\,w}{q\,r}\right)dr\right],\\
\label{10a}
G(r)=\mathrm{sgn}\,k\,(-1)^{n_r}&&\frac{C_{1}^{\pm}}{2}\left[\displaystyle\frac{E-V-m-S}{q(r)}\right]^{1/2}\nonumber\\
&&\times\exp\left[-\int\limits^{r_{0}}_{r}\left(q-\frac{k\,\widetilde{w}}{q\,r}\right)dr\right]. \end{eqnarray}
All integrals in (\ref{10}) and (\ref{10a}) are expressed through the quite complicated combination of the elliptic integrals. But in the cases $\widetilde {E}_r <\widetilde{m}$ and $\widetilde{E}_r>\widetilde{m}$ they can be calculated through elementary functions, using the relations $\sigma/\xi\widetilde{m}^2 \ll1$ and
$\sigma\gamma/\widetilde{E}_r^2 \ll1$ to expand the integrands into power series.

Let us first investigate the asymptotic behavior of the quasiclassical solutions (\ref{10}), (\ref{10a}) at $r\rightarrow 0$ for the lower levels ($\widetilde{E}<\widetilde{m}$) which are basically defined by the Coulomb potential ($\sigma/\xi\widetilde{m}^2 \ll1$).
Note that the larger the Coulomb parameter $\xi$, the smaller is the essential potential $v(r)$ at small distances. Before the evaluation of the asymptotic coefficients $C_{F,G}$ by means of formulas (\ref{10}), (\ref{10a}) it is necessary to expand the quasiclassical momentum $p(r)$ in potential $v(r)$. Then, using the technique of evaluation of the phase integrals from Sec.~\ref{s4} and proceeding in (\ref{10}), (\ref{10a}) to the limit $r\rightarrow 0$, we obtain in zeroth approximation the expressions for the asymptotic coefficients at zero:
\begin{eqnarray}\label{12}
&\displaystyle\left|C_{F}\right|=\sqrt{\frac{\xi}{T\gamma}}\left(\frac{e
\theta_0}{2\gamma^2}\right)^{\gamma}\left[\frac{\theta_0\,(|k|-\gamma)}{\xi(\gamma\widetilde{m}+
|k|\widetilde{E}_0)}\right]^{\frac{\mathrm{sgn}\,k}{2}}\nonumber\\
&\displaystyle\times\left(\frac{\xi\widetilde{E}_0+
\gamma\mu_0}{\theta_0}\right)^{\frac{\xi\widetilde{E}_0}{\mu_0}},\quad\frac{C_{F}}{C_{G}}=\frac{k-\gamma}{\xi}.
\end{eqnarray}
Here
$\theta_0=\sqrt{(\widetilde{E}_0\,k)^2-(\widetilde{m}\,\gamma)^2}$,
and the period of radial oscillations $T$ is given by the previous formula (\ref{8b}).
When we derive the expression (\ref{12}) we use the quasiclassical requirement of the normalization (\ref{t}). Solving the Dirac equation (\ref{1}) at small distances (in range $0<r<c$) one can neglect the term with the linear potential ($\sigma=0$). Having used the asymptotic behavior of the normalized radial functions $F(r)$ and $G(r)$ of the relativistic Coulomb problem \cite{Ahiezer} at $r\rightarrow 0$ and the relations (\ref{9}), we find the more exact (than (\ref{12})) expression for $C_{F}$:
\begin{eqnarray}\label{15}
C_{F}^C=\frac{(2\mu_0)^{\gamma+1/2}}{\Gamma(2\gamma+1)}\left[\frac{(\widetilde{m}+\widetilde{E}_0)
\Gamma(2\gamma+n_r'+1)}{\frac{4\xi\widetilde{m}^2}{\mu_0}\left(\frac{\xi\widetilde{m}}{\mu_0}-k\right)n_r'!}\right]
^{1/2}\nonumber\\
\times\left(\frac{\xi\widetilde{m}}{\mu_0}-k-n_r'\right).
\end{eqnarray}
Here $n_r'=n_r+(1+\mathrm{sgn}\,k)/2$. The formulas (\ref{12}) and (\ref{15}) differ one from another within an error between the Stirling formula
\[n!=\sqrt{2\pi}\exp\left\{(n+1/2)\log n-n\right\}[1+O(n^{-1})],\, n\rightarrow\infty\]
and the $\Gamma$-function.

For states with $\widetilde{E}_r>\widetilde{m}$, when the requirement $\sigma\gamma/\widetilde{E}_r^2\ll 1$ is satisfied, the Coulomb potential is essential only in the range of small distances and can be considered as a small perturbation in the basic range of particle localization (i.e. in classically allowed range $c<r<b$). This gives the possibility to exclude ($\sigma=0$) the linear part of the potential $v(r)$ from the quasiclassical momentum $p(r)$ when evaluating the integrals in exponents (\ref{10}), (\ref{10a}).
Then the asymptotic behavior (at $r\rightarrow 0$) of radial wave functions $F(r)$ and $G(r)$ obtained in this way allows to determine the asymptotic coefficients:
\begin{eqnarray}\label{11}
\left|C_{F}\right|=\sqrt{\frac{\xi}{T\gamma}}\left(\frac{e
\theta}{2\gamma^2}\right)^{\gamma}\left[\frac{\theta\,(|k|-\gamma)}{\xi(\gamma\widetilde{m}
+|k|\widetilde{E}_r)}\right]^{\frac{\mathrm{sgn}\,k}{2}}\nonumber\\
\times\exp\left[\frac{\xi\widetilde{E}_r}{\sqrt{\widetilde{E}_r^2-\widetilde{m}^2}}
\arccos\frac{\xi\widetilde{E}_r}{\theta}\right],
\end{eqnarray}
where quantity $\theta$ is defined in (\ref{root1}), and energy $\widetilde{E}_r$ is given by the formula (\ref{eq5}). Characteristic feature of the considered case is the fact that in the integral (\ref{t}), which defines a period of radial oscillations $T$, only the range of values of the integration variable $r$, where the Coulomb potential can be considered a perturbation, is essential. By neglecting the Coulomb interaction, we arrive at the previous expression (\ref{8a}).

The asymptotic coefficients $A_{F}$, $A_{G}$ of radial wave functions at infinity are important physical parameters of bound states as well. Along with the coefficients $C_{F}$, $C_{G}$ at zero (\ref{9}), the asymptotic coefficients $A_{F,G}$ are continually encountered in quantum mechanics \cite{Landau}, atomic and nuclear physics \cite{Smirnov,Yamabe}, in the converse problem of quantum scattering theory \cite{Newton,Vu} etc. For the potentials (\ref{potential}) the quantities $A_{F,G}$ are related to asymptotic behaviors of the normalized radial wave functions by the relations
\begin{equation}\label{19}
F,G=A_{F,G}\,r^{\widetilde{\gamma}}
\exp\left(-\frac{\sqrt{1-2\lambda}\sigma}{2}\,r^2-
\frac{\eta_1}{\sqrt{1-2\lambda}}\,r\right),
\end{equation}
where $\sigma
r\rightarrow\infty$, $\sigma>0$, $A_{F}=-\sqrt{1-2\lambda}\,A_{G}$,
$\widetilde{\gamma}=\displaystyle\frac{\eta_2^2}{2(1-2\lambda)^{3/2}\sigma}-
\frac{\lambda\xi}{\sqrt{1-2\lambda}}$, and the parameter $\lambda$
has values in the range $0\leqslant\lambda<1/2$.

In the below-barrier range $r>r_1=b$ far from the turning point $r_1=b$ under the requirements $\sigma/\xi\widetilde{m}^2\ll1$ and $\widetilde{E}<\widetilde{m}$, after the evaluation of integrals the quasiclassical solutions (\ref{2e}) are of the form of decreasing exponents
\begin{widetext}
\begin{eqnarray}\label{13}
&&\left(\begin{array}{ll}F\\G\end{array}\right)\approx\frac{1}{\sqrt{T
q_0}}\left(\begin{array}{ll} \sqrt{\widetilde{m}+\widetilde{E}_0+
(1-2\lambda)\sigma r}\\-\sqrt{\widetilde{m}-\widetilde{E}_0+\sigma
r}\end{array}\right)\left(\frac{4\mu_0^4\theta_0^{-1}
r}{\mu_0^2+\mu_0 q_0+\eta_{10}\sigma r}\right)^
{\frac{\xi\widetilde{E}_0}{\mu_0}}\left(\frac{\xi\widetilde{m}-k\mu_0}
{\xi\widetilde{m}+k\mu_0}\right)^{1/4}\left(\frac{\xi\widetilde{E}_0-\gamma\mu_0}
{\xi\widetilde{E}_0+\gamma\mu_0}\right)^{\gamma/2}\nonumber\\
&&\times\left(\frac{\sqrt{1-2\lambda}q_0+
(1-2\lambda)\sigma
r+\eta_1+\xi\widetilde{E}(1-2\lambda)\sigma/\mu^2}{\sqrt{1-2\lambda}[\mu+
\xi(\lambda\mu^2+2\eta_1\widetilde{E})/\mu^3]
+\eta_1+\xi\widetilde{E}(1-2\lambda)\sigma/\mu^2}\right)^{\widetilde{\gamma}}\exp\left[-\frac{q_0
r}{2}+\frac{\eta_1(\mu-q_0)}{2(1-2\lambda)\sigma}+
\frac{\xi\widetilde{E}(\mu+q_0)}{2\mu^2}\right],
\end{eqnarray}
\end{widetext}
where $q_0=\sigma\sqrt{(1-2\lambda)(r-c)(r-d)}$. The estimates show that by the satisfaction of the requirements $\sigma/\xi\widetilde{m}^2\ll 1$ and $\widetilde {E}<\widetilde{m}$ there is quite long range of distances $r$ which are much larger than size of the Coulomb hydrogen-like system ($r\gg\langle r\rangle$, see (\ref{8b}) or (\ref{8c}) and much smaller than the distance $\widetilde{r}\approx(\widetilde{E}\xi/\eta_1\sigma)^{1/2}$ at which the Coulomb interaction becomes quantitatively comparable with the long-range interaction. In this range as the wave functions of zeroth approximation it is natural to take the unperturbed radial functions $F$ and $G$ of the relativistic Coulomb problem, and the potential $v(r)$ can be considered as a small perturbation.
Neglecting it, we arrive at the following quasiclassical expressions for $F$ and $G$
\begin{eqnarray}\label{13a}
&&\left(\begin{array}{ll}F\\G\end{array}\right)=\left(\begin{array}{ll}
\sqrt{\widetilde{m}+\widetilde{E}_0}\\-\sqrt{\widetilde{m}-\widetilde{E}_0}
\end{array}\right)A^{\mbox{\footnotesize WKB(as)}}_Cr^
{\frac{\xi\widetilde{E}_0}{\mu_0}}e^{-\mu_0 r}
\nonumber\\
&&=\displaystyle\frac{1}{\sqrt{T \mu_0}}\left(\begin{array}{ll}
\sqrt{\widetilde{m}+\widetilde{E}_0}\\-\sqrt{\widetilde{m}-\widetilde{E}_0}
\end{array}\right)\left(\frac{\xi\widetilde{m}-k\mu_0}
{\xi\widetilde{m}+k\mu_0}\right)^{1/4}\nonumber\\
&&\times\left(\frac{\xi\widetilde{E}_0-\gamma\mu_0}
{\xi\widetilde{E}_0+\gamma\mu_0}\right)^{\gamma/2}\left(\frac{2\mu_0^2
r}{\theta_0}\right)^{\frac{\xi\widetilde{E}_0}{\mu_0}}e^{-\mu_0
r}.
\end{eqnarray}
Equating (\ref{13a}) to the asymptotic (at $r\rightarrow\infty$) representation of solutions of the Dirac equation in the Coulomb field \cite{Ahiezer}
\begin{eqnarray}\label{14}
\left(\begin{array}{ll}F\\G\end{array}\right)=
\left(\begin{array}{ll}
\sqrt{\widetilde{m}+\widetilde{E}_0}\\-\sqrt{\widetilde{m}-\widetilde{E}_0}
\end{array}\right)A_Cr^{\frac{\xi\widetilde{E}_0}{\mu_0}}e^{-\mu_0 r}
\end{eqnarray}
we obtain the explicit expression for the period of radial oscillations of the classical relativistic particle:
\begin{eqnarray}\label{17}
T=\frac{1}{2 \mu_0 |A_C|^2}\left(\frac{\xi\widetilde{m}-k \mu_0}{\xi\widetilde{m}+k
\mu_0}\right)^{1/2}\left(\frac{2e\mu_0^2}{\theta_0}\right)^{\frac{2\xi\widetilde{E}_0}{\mu_0}}\nonumber\\
\times\left(\frac{\xi\widetilde{E}_0-\gamma\mu_0}{\xi\widetilde{E}_0+\gamma\mu_0}\right)^{\gamma}.
\end{eqnarray}
Here $A_C$ is the asymptotic coefficient of the Dirac radial wave functions in the Coulomb potential:
\begin{equation}\label{18} |A_C|=\left[\frac{(\xi\,\widetilde{m}-k\,\mu_0)\,\mu_0}{2\,\xi\,\widetilde{m}^2\,
\Gamma\left(2\gamma+n_r'+1\right)n_r'!}\right]^{1/2}
(2\mu_0)^{\frac{\xi\widetilde{E}_0}{\mu_0}}.
\end{equation}
Comparison of (\ref{13a}) and (\ref{14}) shows that their exponential and power factors are the same, however, the asymptotic coefficients $A^{\mbox{\footnotesize WKB(as)}}_C$ and $A_C$ differ within an error between the Stirling formula and $\Gamma$-function.

The formula (\ref{13a}) is obtained by neglecting the linear part of the potential $v(r)$. This approximation was argued above by means of the circumstance that under the quasiclassical requirements ($\sigma/\xi\widetilde{m}^2 \ll1$) there is a range of distances $\langle r\rangle\ll r\ll\tilde{r}$, in which the distortion of a wave function, caused by action of the linear part of the potential $v(r)$, can still be neglected and there is the law of decreasing radial wave functions (\ref{13a}) that is characteristic for the relativistic Coulomb problem. Change of the law of decreasement (\ref{13a}) of functions $F(r)$ and $G(r)$ to (\ref{19}) at $r\gg\tilde{r}$ appears because in EP $U(r,E)$ we have taken into account the quadratic (in $\sigma r$) terms which increase with increasing $r$ more rapidly than others and so play a role of the perturbation which destroys the asymptotic regime (\ref{13a}).  As a result of such an account, using the quasiclassical approximation (\ref{13}) for the normalized radial wave functions $F$ and $G$ at large $r$, we obtain the following expression for the asymptotic coefficient at infinity
\begin{align}\label{19b}
 A_{F}=2\mu_0 A_C(1-2\lambda)^{\widetilde{\gamma}+1/4}
\left(\frac{\sqrt{1-2\lambda}\,\mu_0+\eta_{10}}{2}\right)^{-\frac{\xi\widetilde{E}_0}{\mu_0}-\widetilde{\gamma}}\nonumber\\
\times\sigma^{\widetilde{\gamma}}
\left(\frac{\mu_0^2}{\sigma}\right)^{\frac{\xi\widetilde{E}_0}{\mu_0}}
\exp\left[-\frac{(\sqrt{1-2\lambda}\,\mu-\eta_1)^2}{4(1-2\lambda)^{3/2}\sigma}\right].
\end{align}

We now proceed to the other limiting case $\sigma\gamma/\widetilde {E}^2\ll 1$ when the centrifugal potential $\gamma^2/2mr^2$ does not play an essential role at large distances and can be omitted in the quasiclassical momentum $p(r)=iq(r)$. Having expanded the quantity $q(r)=|p(r)|$ in powers of the Coulomb potential and calculated the integrals in exponents in (\ref{2e}), under the requirement $\sigma\gamma/\widetilde{E}^2\ll1$, in the asymptotic domain $r\rightarrow\infty$ we arrive at formulas of type of (\ref{19}) for $F$ and $G$, in which
\begin{align}\label{19a}
 A_{F}=\frac{(1-2\lambda)^{1/4}}{\sqrt{T}}\left(\frac{2(1-2\lambda)\sigma}
{\eta_2}\right)^{\widetilde{\gamma}}
\exp\left[-\frac{2\eta_1^2-\eta_2^2}{4(1-2\lambda)^{3/2}\sigma}\right.\nonumber\\
+\left.\frac{\xi\widetilde{m}\eta_2}{2\sqrt{1-2\lambda}(\widetilde{E}^2-\widetilde{m}^2)}
+\frac{\xi\widetilde{E}}{\sqrt{\widetilde{E}^2-\widetilde{m}^2}}\arccos
\left(-\frac{\eta_1}{\eta_2}\right)\right],
\end{align}
and the period $T$ is determined by the previous expression (\ref{8a}).

\section{Energy spectrum of the massless fermion in the external scalar field of the funnel type}\label{s11}

In this section, we shall study energy spectrum of the Dirac equation (\ref{1}) for massless fermion ($m=0$) in the external scalar field of the form
\begin{equation}\label{58}
 S(r)=-\frac{\xi'}{r}+\sigma r,\quad \sigma>0; \quad V(r)=0.
\end{equation}
Physical details of the considered model (\ref {58}) can be found in \cite{Lazur}.

For the potential (\ref{58}) and particle with zero mass the quasiclassical quantization condition (\ref{3a}) becomes the transcendental equation
\begin{align}
&\frac{E^2+2\sigma(\xi'-\gamma)}{4\sigma}-\frac{k}{\sigma (b+
a)\pi}\left[2b\left(\frac{\Pi(\nu^*_+,\chi^*)}{b^2-P_+^2}\right.\right.\nonumber\\
&\left.\left.+
\frac{\Pi(\nu^*_-,\chi^*)}{b^2-P_-^2}\right)-\left(\frac{1}{b+P_+}+\frac{1}{b+P_-}\right)K(\chi^*)\right]=
n_r+\frac{1}{2}.\label{66}
\end{align}
Here $K(\chi^*)$ and $\Pi(\nu^*,\chi^*)$ are the complete elliptic integrals of the first and third kind, and the following notations are used:
\begin{eqnarray*}
&\chi^*=\sqrt{\displaystyle \frac{
E^2+2\sigma(\xi'-\gamma)}{E^2+2\sigma(\xi'+\gamma)}},\quad
\nu^*_{\pm}=\chi^*\,\displaystyle
\frac{P_{\pm}+r_0}{P_{\pm}-r_0},&\\
&P_{\pm}=\frac{1}{2\,\sigma}\left(-E\pm\sqrt{E^2+4\xi'\sigma}\right).&
\end{eqnarray*}
where the turning points $r_0=b$ and $r_1=a$ are determined by the equations
\[
  b,a=\frac{1}{2^{1/2}\sigma}\sqrt{E^2+2\xi'\sigma \mp
  \sqrt{(E^2+2\xi'\sigma)^2-4\,\sigma^2\gamma^2}},
\]
$\gamma=\sqrt{k^2+\xi'^2}$. In the case $ \sigma\rightarrow 0$ the equation (\ref {66}) can be solved in an explicit form. At small values of parameter $\sigma$ (namely at $ \sigma\lesssim 0.2 \, \mbox {GeV} ^2$) the condition $E^2 _ {n_r k} \gg \sigma\gamma $ for all possible values of level energy $E _ {n_r k} $ is satisfied very well. In this case the formula given above become appreciably  simpler and the equation (\ref {66}) for a quasiclassical spectrum ultimately assumes the rather simple form
\begin{eqnarray}\label{67}
\frac{E_{n_r k}^2}{2\sigma}=N'+\frac{\sigma k}{E_{n_r
k}^2}\left[\frac{1}{\pi}\left(\log\frac{4E_{n_r
k}^2}{\sigma\gamma}-1\right)-R\right]\nonumber\\
+O\left(\left(\frac{\sigma
\gamma}{E_{n_r k}^2}\right)^2\right),
\end{eqnarray}
where $\sigma>0$ and
\begin{eqnarray*}
&\displaystyle N'=2n_r+1+\gamma-\xi'+2B\,\mbox{sgn}\,k,&\\
&\displaystyle B=\frac{1}{\pi}\arctan\sqrt{\frac{\gamma+\xi'}{\gamma-\xi'}},\,
R=\frac{2\xi'}{k^2}\left(\frac{\xi'}{\pi}+\frac{2\gamma^2B}{|k|}\right).&
\end{eqnarray*}
In that case, when the Coulomb-like term in the potential (\ref {58}) is absent ($ \xi ' =0$), the equation (\ref {67}) coincides exactly with the quasiclassical quantization rule for energy levels in the scalar well $U (r, E)$ generated by linear confining interaction \cite{Simonov1}
$S(r)=\sigma r$, $V(r)=0$.

The equation (\ref{67}) for $E_{n_r k}$ is easy for solving numerically. Comparison of results of such calculations $E_{n_r k}$ with exact values \cite {Mur4} (see tab.~\ref{tab11}), obtained by numerical integration of the Dirac system (\ref {1}), shows that the quasiclassical equation (\ref{67}) provides acceptable accuracy of calculation of the energy spectrum: even for the lower states with $n_r\sim 1$ the error (\ref{67}) in determination of $E_{n_r k}$ does not exceed $5\%$ and rapidly decreases with increasing $n_r$.
\begin{table} \caption{{\footnotesize The eigenvalues $E_{n_r k}$ of the massless Dirac equation with the scalar potential (\ref{58}) for
$\xi=0.4$, $\sigma=0.18\,\mbox{GeV}^2$}}\label{tab11}
\begin{center}
\begin{tabular}{|c|c|c|c|c|c|}
 \hline
\multicolumn{2}{|c|}{\mbox{States}}&$E^{\mbox{{\footnotesize
num}}}_{n_r k}$, GeV& $E^{\mbox{{\footnotesize WKB}}}_{n_r k}$,
GeV&$E^{\mbox{{\footnotesize as}}}_{n_r k}$, GeV&
$E^{\mbox{{\footnotesize as}}}_{n_r k}$, GeV\\
\cline{1-2}
$n_r$&$k$&(\ref{1})&(\ref{66})&(\ref{67})&(\ref{68})\\
\hline0&-1&0.5568&0.5581&0.6530&0.6568\\
1&-1&1.0293&1.0298&1.0489&1.0489\\
2&-1&1.3379&1.3382&1.3471&1.3470\\
3&-1&1.5862&1.5863&1.5917&1.5917\\
4&-1&1.7999&1.8000&1.8037&1.8037\\
\hline0&-2&0.8217&0.8220&0.8561&0.8554\\
1&-2&1.1898&1.1899&1.2003&1.2001\\
2&-2&1.4649&1.4650&1.4704&1.4703\\
3&-2&1.6949&1.6949&1.6983&1.6983\\
4&-2&1.8966&1.8966&1.8990&1.8990\\
\hline0&1&0.9340&0.9335&0.9054&0.9055\\
1&1&1.2568&1.2566&1.2457&1.2457\\
2&1&1.5143&1.5141&1.5079&1.5097\\
3&1&1.7346&1.7345&1.7304&1.7304\\
4&1&1.9302&1.9302&1.9272&1.9272\\
\hline0&2&1.0996&1.0994&1.0857&1.0853\\
1&2&1.3846&1.3845&1.3781&1.3780\\
2&2&1.6217&1.6216&1.6177&1.6177\\
3&2&1.8289&1.8288&1.8261&1.8261\\
4&2&2.0152&2.0152&2.0132&2.0132\\
\hline
\end{tabular}
\end{center}
\end{table}
Along with the direct numerical solution of transcendental equations (\ref{66}) and (\ref{67}) it is worthwhile to construct (by means of some simplifications or approximations) the approximate analytical expressions for level energy that would allow without difficulty to trace dependence of $E_{n_r k}$ on quantum numbers $n_r$, $k$ and parameters of interaction model (\ref {58}). Solving (\ref{67}) by the method of iterations we arrive at the analytical expression for energy
\begin{eqnarray}
&&\varepsilon_{n_r k}=\frac{E_{n_r,k}}{\sqrt{\sigma}}=
\pm\left\{N'-\xi'+\left[(N'-\xi')^2\right.\right.\nonumber\\
&&\left.\left.+2k\left(\frac{1}{\pi}\left(\log
\frac{8(N'-\xi')}{\gamma}-1\right)-R\right)\right]^{1/2}\right\}.\label{68}
\end{eqnarray}
The positive sign of the root corresponds to energy of a particle, and the negative one corresponds to the antiparticle energy  taken with a minus sign.

\section{Summary}

The most important results of investigation performed can be summarized as follows:
\begin{enumerate}
\item{The relativistic potential quark model of $Q\bar{q}$-mesons in which the light quark motion is described by the Dirac equation with a scalar-vector interaction and the heavy quark is considered a local source of the gluon field is constructed. The effective interquark interaction is described by a combination of the perturbative one-gluon exchange potential $V_{\mathrm{Coul}}(r)=-\xi/r$ and the long-range Lorentz-scalar and Lorentz-vector linear potentials $S_{\mathrm{l.r.}}(r)=(1-\lambda)(\sigma r+V_0)$ and $V_{\mathrm{l.r.}}(r)=\lambda(\sigma r+V_0)$. It is established that the quark confinement always arises when the Lorentz-scalar part $S_{\mathrm{l.r.}}$ of the long-range interquark interaction prevails over the Lorentz-vector one $V_{\mathrm{l.r.}}$.}

\item{The new asymptotic method of calculation of quantization integrals that based on splitting an integration interval into two segments in each of which only the dominating interaction type is taken into account exactly while the other interaction types are treated as perturbations is elaborated. Approximative analytical expressions for energy spectrum of heavy-light mesons obtained within quasiclassical approach at $\sigma\gamma/\widetilde{E}^2\ll 1$ are asymptotically exact in the limit $n_r\rightarrow\infty$ and ensure a high accuracy of calculations even for states with the radial quantum number $n_r\sim 1$.}

\item{In the framework of the considered model we have obtained the satisfactory description of the mass spectrum of $D$-, $D_{s}$-, $B$-, and $B_{s}$-mesons. We show that the fine structure of P-wave states in heavy-light mesons is primarily sensitive to the choice of two parameters: the strong-coupling constant $\alpha_s$ and the coefficient $\lambda$ of mixing of the long-range scalar and vector potentials $S_{\mathrm{l.r.}}(r)$ and $V_{\mathrm{l.r.}}(r)$. The best agreement between the theoretical predictions and experimental data exists when the mixing coefficient $\lambda=0.3$.}

\item{Using WKB method the convenient analytical formulas for asymptotic coefficients of wave function at zero and infinity and mean radii of the $Q\bar{q}$ mesons.}

\end{enumerate}

\section*{Appendix}

We consider the quantization integral $J_1$. We rewrite the expression for $J_1$ in (\ref{A4}) as the sum of
integrals
$$
J_{1}=-|\sigma|\,\sqrt{1-2\lambda}\,(l\Im_{-1}+h\Im_{0}+g\Im_{1}+f\Im_{2}+\Im_{3}),
$$
$$\Im_{n}=\displaystyle\int\limits_{b}^{a}\frac{r^n}{R}\,dr,\qquad n= -1, 0, 1, 2, 3,\ldots,$$
where the quantities $f$, $g$, $h$, and $l$ are determined in (\ref{A2}) and the quantity $R(r)$ is determined in (\ref{eq4a}). After the standard change of the integration variable \cite{Beitmen}
$$
r=\frac{b(a-c)-c(a-b)\sin^{2}\varphi}{a-c-(a-b)\sin^{2}\varphi}
$$
the integrals $\Im_{n}$ are expressed in terms of the complete elliptic integrals of the first, second, and third kind, which are written in the conventional notation \cite{Prudnikov} as
$$\begin{array}{c}
\displaystyle
F(\chi)=\int\limits^{\pi/2}_{0}\frac{d\varphi}{\triangle},\quad
E(\chi)=\int\limits^{\pi/2}_{0}\triangle\,d\varphi, \\
\displaystyle\Pi(\nu,\chi)=\int\limits^{\pi/2}_{0}\frac{d\varphi}{(1-\nu
\sin^2\varphi)\triangle},
\end{array}\eqno(\mbox{A}.1)
$$
$$\displaystyle
\triangle=\sqrt{1-\chi^{2}\sin^{2}\varphi},\quad \nu=
\displaystyle\frac{a-b}{a-c},\quad \chi=\sqrt{\nu
\frac{(c-d)}{(b-d)}}.$$

We thus obtain the representations for $\Im_{-1},\ldots, \Im_{3}$:
$$
\Im_{-1}=\displaystyle\int\limits_{b}^{a}\frac{dr}{r\,R}=\frac{2}{\sqrt{\left(a-c\right)\left(b-d\right)}\,bc}\Bigl[bF
\left(\chi\right)\Bigr.$$
$$\hspace{-20mm}-\left.\left(b-c\right)\Pi\left(\frac{c}{b}\nu,\chi\right)\right],
\eqno(\mbox{A}.2)$$
$$\Im_{0}=\displaystyle\int\limits_{b}^{a}\frac{dr}{R}=\frac{2}{\sqrt{\left(a-c\right)\left(b-d\right)}}\,F
\left(\chi\right),\eqno(\mbox{A}.3)$$
$$\Im_{1}=\displaystyle\int\limits_{b}^{a}\frac{r\,dr}{R}=\frac{2}{\sqrt{\left(a-c\right)\left(b-d\right)}}\left[cF
\left(\chi\right)\right.$$
$$\hspace{-25mm}\left.+\left(b-c\right)\Pi\left(\nu,\chi\right)\right],\eqno(\mbox{A}.4)$$
$$\hspace{-8mm}\Im_{2}=\int\limits_b^a
\frac{r^2\,dr}{R}=\frac{2}{\sqrt{\left(a-c\right)\left(b-d\right)}}\Bigl[c^{2}F
\left(\chi\right)\Bigr.$$
$$\hspace{12mm}+c\left(b-c
\right)\Pi\left(\nu,\chi\right)+\left.\left(b-c\right)^{2}T_{2}\left(\textstyle{\pi\over2},\nu,\chi\right)\right],\eqno(\mbox{A}.5)$$
$$\hspace{-8mm}\Im_{3}=\int\limits_b^a
\frac{r^3\,dr}{R}=\frac{2}{\sqrt{\left(a-c\right)\left(b-d\right)}}\Bigl[c^{3}F
\left(\chi\right)\Bigr.$$
$$\hspace{5mm}+3c^{2}\left(b-c\right)\Pi\left(\nu,\chi\right)+3c\left(b-c\right)^{2}T_{2}\left(\textstyle{\pi\over2},\nu,\chi\right)$$
$$\hspace{-27mm}\left.
+\left(b-c\right)^{3}T_{3}\left(\textstyle{\pi\over2},\nu,\chi\right)\right].\eqno(\mbox{A}.6)$$

The integrals of the form
\[T_{n}\left(\varphi,\nu,\chi\right)=\displaystyle\int\limits_{0}^{\varphi}
\frac{d\varphi}{\left(1-\nu\sin^{2}\varphi\right)^n\triangle}\]
are calculated using the recurrence relation
$$
T_{n-3}=\frac{1}{(2n-5)\chi^2}\left\{\frac{-\nu^2\sin\varphi\cos\varphi\triangle}
{\left(1-\nu\sin^{2}\varphi\right)^{n-1}}\right.$$
$$+2(n-2)\left[3\chi^2-\nu(1+\chi^2)\right]T_{n-2}$$
$$-(2n-3)\left[\chi^{2}\left(3-2\nu\right)+\nu\left(\nu-2\right)\right]T_{n-1}$$
$$\Biggl.
+2(n-1)(\chi^2-\nu)(1-\nu)T_{n}\Biggr\}.$$

We analogously find the integrals in the expression for $J_2$ in (\ref{A4}):
$$
\displaystyle\int\limits_{b}^{a}\frac{dr}{(r-\lambda_{\pm})\,R}=\frac{2}{\sqrt{\left(a-c\right)\left(b-d\right)}\,
(b-\lambda_{\pm})(\lambda_{\pm}-c)}$$
$$\times\left[(\lambda_{\pm}-b)F\left(\chi\right)-\left(b-c
\right)\Pi\left(\frac{(\lambda_{\pm}-c)}{(\lambda_{\pm}-b)}\nu,\chi\right)\right],
\eqno(\mbox{A}.7)$$

After expressions (A.2)--(A.7) are substituted in integrals (\ref{A4}), quantization condition (\ref{3a}) becomes
transcendental equation (\ref{5}), where
\begin{eqnarray}
\nu_{\pm}&=& \frac{\lambda_{\pm}-c}{\lambda_{\pm}-b}\,\nu, \quad
\Re=\left(1-\nu\right)\left(\chi^{2}-\nu\right), \nonumber\\
N_{1}&=&\frac{\chi^{2}\left(b-c\right)}{4}- \frac{3\aleph
\left(b-c\right)}{8\left(1-\nu\right)}-\frac{\left(\chi^{2}-\nu\right)}{2}
\left(f+3c\right)\nonumber\\
&+&\frac{\Re}{\left(b-c\right)^{2}}\left(c^{3}+c^{2}f+cg+h+l/c\right),\nonumber\\
\aleph&=&\chi^2\left(3-2\nu\right)+\nu(\nu-2), \nonumber\\
N_{2}&=&-\frac{\nu}{2}\left[f+3c+\frac{3}{4}\frac{\left(b-c\right)\aleph}
{\Re}\right],\nonumber\\
N_{3}&=&\frac{1}{2}\left[\frac{3}{4}\frac{\left(b-c\right)\aleph^{2}}{\Re}+\frac{2\Re}{\left(b-c\right)}
\left(3c^{2}+2cf+g\right)\right. \nonumber\\
&+&\Bigl.\left(b-c\right)\left(\left(1+\chi^{2}\right)\nu-3\chi^{2}\right)+\aleph\left(f+3c\right)\Bigr],\nonumber\\ N_{4}&=&-\frac{\Re}{\left(b-c\right)}\frac{l}{b c},\quad
N_{5}=[(b-\lambda_{+})(\lambda_{+}-c)]^{-1}, \nonumber\\
N_{6}&=&[(b-\lambda_{-})(\lambda_{-}-c)]^{-1}, \quad \nonumber\\
N_{7}&=&\frac{2}{(\lambda_{+}-c)
(\lambda_{-}-c)}\left(c+\frac{\widetilde{E}+\widetilde{m}}{2(1-2\lambda)\sigma}\right).\nonumber
\end{eqnarray}

We analogously find the integrals appearing when calculating the mean radii by formula (\ref{6}). We
present the quantities $n_i$ ($i=1,\ldots, 6$) in formulas (\ref{7}) and (\ref{7a}):
\begin{align*}
&n _1=
\widetilde{E}\left(c^2-\frac{\left(b-c\right)^2}{2\left(1-\nu\right)}\right)-\lambda
\sigma\left(c^3-\frac{3c\left(b-c\right)^2}{2\left(1-\nu\right)}\right.\\
&+\left.\frac{\left(b-c\right)^3}{4\Re}\left(\chi^2-\frac{3\aleph}{2\left(1-\nu\right)}
\right) \right)+\xi c,
\end{align*}
\begin{align*}
n_2&=-\frac{\nu\left(b-c\right)^2}{2\Re}\left[\widetilde{E}-3\lambda
\sigma
\left(c+\frac{\left(b-c\right)\aleph}{4\Re}\right)\right],
\end{align*}
\begin{align*}
&n_3=\left(b-c\right)\left[\widetilde{E}\left(2c+\frac{\left(b-c\right)\aleph}{2\Re}\right)\right.\\
&- \lambda  \sigma
\left(3c^2+\frac{\left(b-c\right)\aleph}{2\Re}\Biggl
(3c-\Biggr.\right.\\
&-\left.\left.\left.\frac{\left(b-c\right)\left(3\chi^2-\nu
\left(1+\chi^2\right)\right)}{\aleph}+\frac{3\left(b-c\right)\aleph}{4\Re}\right)\right)
+\xi\right],
\end{align*}
\[n_4=c \widetilde{E}-\lambda  \sigma
\left(c^2-\frac{\left(b-c\right)^2 }
{2\left(1-\nu\right)}\right)+\xi,\,n_5=\frac{\lambda  \sigma \nu \left(b-c\right)^2}{2\Re},\]
\[n_6=\left(b-c\right)\left[\widetilde{E}-\lambda  \sigma
\left(2c+\frac{\left(b-c\right)\aleph} {2\Re}\right)\right].\]

\end{document}